\PassOptionsToPackage{square}{natbib}

\documentclass[onlinecite,showpacs,aps,twocolumn,amsmath,amsfonts,floatfix,showpax]{revtex4-1}

\usepackage{epsf}
\usepackage{graphicx}           
\usepackage{color}
\usepackage{amsmath}
\usepackage{revsymb4-1}
\usepackage{rotating}
\usepackage{tabulary}
\usepackage{setspace}
\usepackage{multirow}
\usepackage{color}
\usepackage{wasysym}
\definecolor{green0}{rgb}{0,0.5,0}
\usepackage{bm}
\usepackage{fixmath}
\usepackage{amsbsy}
\usepackage{pifont}
\usepackage{diagbox}

\begin{document}

\title{Entanglement structure of the Hubbard model in momentum space}
\onecolumngrid

\author{G.\ Ehlers$^1$, J. S\'olyom$^2$, \"O.~Legeza$^{2}$, R. M. Noack$^1$} 
\affiliation{
   $^1$ Fachbereich Physik, Philipps-Universit\"at Marburg, 35032 Marburg, Germany\\
   $^2$ Strongly Correlated Systems ``Lend\"ulet'' Research group,
   Wigner Research Centre for Physics, H-1525 Budapest, Hungary} 

\date{\today}

\begin{abstract}
We study the properties of the ground states of the one- and
two-dimensional Hubbard models at half-filling and moderate doping
using entanglement-based measures, 
which we calculate numerically using the momentum-space
density matrix renormalization group (DMRG).  
In particular, we investigate quantities such as the
single-site entropy and two-site mutual information of single-particle
momentum states as well as the behavior of the bipartite subsystem
entropy for partitions in momentum space.
The distribution of these quantities in momentum space 
gives insight into the fundamental nature of the ground state, 
insight that can be used to make contact with
weak-coupling-based analytic approaches and to optimize numerical
methods, the momentum-space DMRG in particular.  We study the site and
subsystem entropies as a function of interaction strength $U$ and
system size.  In both the one-
and two-dimensional cases, we find that 
the subsystem entropy scales proportionally to $U^2$ for weak $U$ and
proportionally to volume.  Nevertheless, the optimized momentum-space
DMRG can provide variationally accurate results for the
two-dimensional Hubbard model at weak coupling for moderate system
sizes.
\end{abstract}


\maketitle

\section{Introduction}

Simulating quantum systems on classical computers is a major challenge
in theoretical physics, a challenge that has been addressed 
by a number of recent algorithmic 
developments~\cite{Vidal-2003a,LeBlanc-2015}.
Such simulation is particularly difficult when strong correlations,
as reflected in a high level of 
entanglement, are present~\cite{Amico-2008}.
For one-dimensional
systems, the matrix-product-state--based 
(MPS--based)~\cite{Ostlund-2005,Verstaete-2004} 
density matrix renormalization group (DMRG) method~\cite{White-1992} 
is the most effective tool to calculate the properties of the ground
and selected excited states.
In higher dimensions, the DMRG is less efficient at representing the
entanglement structure of states due to its one-dimensional topology. 
Recent developments have led to several alternative methods 
that potentially overcome the limitations of the DMRG. 
These include, tensor network states (TNS)~\cite{Nishino-2001},
the multi-scale entanglement renormalization (MERA)~\cite{MERA},
the projected entangled pair states (PEPS)~\cite{PEPS},
the tree-tensor network state (TTNS)~\cite{TTNS},
quantum Monte Carlo 
(QMC)~\cite{QMC-White,QMC-S_Zhang,AFQMC,QMC-Becca,QMC-Tocchio},
and the density matrix embedding theory (DMET)~\cite{DMET-1,DMET-2}.
All such methods have 
advantages and disadvantages; each of these methods has 
competitive performance in particular circumstances.
However, there is no single method yet that can be applied efficiently
as a black-box tool to a general quantum lattice problem.
Therefore, despite substantial effort over many years, 
accessible system sizes in more than one dimension
are still strongly limited. 

For example, in two and higher dimensions, QMC is a
potentially very powerful method but it suffers from the so-called 
fermion
sign problem~\cite{QMC-Loh,QMC-Troyer,AFQMC}, which
occurs when fermionic systems and spin
systems with frustration are treated.
The domain of application of PEPS is limited due to the 
expensive scaling of the computational cost with bond dimension when
highly entangled systems are treated~\cite{PEPS,Orus-2014}.
In addition, when the infinite-system version of PEPS, iPEPS, is used,
the proper choice of the size of the unit  
cell can significantly effect the final solution~\cite{Corboz-2014}.
Another promising method for simulating two-dimensional systems 
is the DMET, a method that is based on the numerically exact
calculation of the ground state (or other particular state) of a
finite-size cluster embedded to an environment via a density matrix.
The cluster calculation is typically carried out using exact
diagonalization or the DMRG, 
whereas the degrees of freedom of the environment are approximated.
The accuracy therefore depends on the level of truncation of the
environment degrees of freedom.
Within the DMRG, a two-dimensional system must be mapped to a
one-dimensional lattice topology due to the nature of the MPS, which
must then be optimized by sweeping through the system in a
``snake-like'' manner~\cite{Noack-White}.
Thus, local interactions become  
nonlocal with respect to the MPS, and an efficient representation of
a locally entangled two-dimensional state becomes 
impossible.
An overview of recent state-of-the-art results on the
two-dimensional Hubbard model 
calculated using a variety of numerical algorithms has been
given in a recent review~\cite{LeBlanc-2015}.
One interesting conclusion is that,
in spite of the limitations of MPS--based methods for
two-dimensional systems, 
competitively accurate results have been obtained for
two-dimensional frustrated spin systems using the DMRG.
For example, for the Heisenberg model on the kagome lattice,
DMRG calculations yield results that have among the best levels of
variational accuracy~\cite{White-kagome,Schollwock-kagome}
and are in excellent agreement with results from
MERA~\cite{Vidal-kagome} and PEPS~\cite{PEPS-kagome}.

The amount of entanglement present when partitioning a system,
however, is basis dependent~\cite{Legeza-2003c,PEPS,Szalay-2015}.
Therefore, a given problem could possibly
be solved with significantly less computational cost if an 
optimal representation could be found~\cite{Fertitta-2014,Krumnow-2015}.
For example, the DMRG treatment 
of a two-dimensional noninteracting fermionic problem is
exponentially difficult in real space due to the high level of
entanglement, whereas in    
a momentum-space representation the solution is a simple product 
state, which can be represented exactly with a low-dimensional
MPS~\cite{Legeza-2003c}.
For the case of non-vanishing 
interactions, however, interactions that are
local in a particular basis (typically real space) become 
nonlocal after a change of basis, influencing the
scaling behavior of entanglement with system size dramatically.
For systems with sufficient locality, i.e., with hopping and
interaction terms that have a finite range within a lattice of a
particular dimensionality, the scaling of the entropy for ground
and certain low-lying excited states is thought to obey the entropy
area law~\cite{Eisert-2010}.
Thus, such models with short-range interactions can, generically, be
simulated efficiently using tensor network state algorithms.
When locality is lost, the area law, in general, can break
down, making efficient treatment of the problem difficult~\cite{Legeza-2006}.
Nevertheless, for 
small interaction strengths and for a given error threshold, it is
still potentially possible to find a representation that is
computationally more efficient for particular system sizes. 
In addition, the weak-coupling regime is often well approximated 
by analytical approaches~\cite{Solyom}, providing 
connections between numerical and analytical results.
Furthermore, geometrical aspects of a Hamiltonian in
a particular representation can have
a more compact description in another representation; in particular,
what is long range in real space tends to be short range in momentum
space and vice versa.

Motivated by the arguments above, here we study and compare the
real-space and momentum-space representations of the Hubbard model in
one and two dimensions from the point of view of quantum information
entropies.
Although previous numerical studies of the Hubbard model 
using the momentum-space DMRG method (k-DMRG) 
date from
as long as  two decades ago~\cite{Xiang-1996,Nishimoto-2002,Legeza-2003c},
these studies were severely limited in the number
of block states kept,  
worked with non-optimally
ordered MPSs for the two-dimensional system, and did not study
the behavior of quantum-information-based quantities.
Therefore, a rigorous analysis of the scaling of the entanglement
entropy and other measures of quantum correlation as a function of system
sizes and interaction strengths could not be carried out.
In this paper, we reexamine the Hubbard model using the k-DMRG,
presenting state-of-the-art results using
a modern MPS--based DMRG code that optimizes the  mapping of the
lattice to an
MPS and can keep up to of the order of $50\,000$ block
states for the momentum-space Hubbard model.
To study the intrinsic properties of the MPS representation,
we carry out a detailed analysis of entanglement scaling
with system size and interaction strength $U$ in one and two
dimensions. 

Our study is also motivated by the fact that information on the
dimensionality and geometry of the lattice as well as range of the
hopping is encoded in the kinetic term, which is diagonal in momentum
space.
Thus, the k-DMRG is less sensitive
than the real-space representation to changes in these aspects of the
model, that is, more complex cases that add computational cost in
real space could potentially be treated with little additional cost in
momentum space; it is important to determine to what extent this is
the case.
Furthermore, the momentum-space representation is intrinsically
translationally invariant, so that 
finite-size corrections can be treated by scaling systems with periodic
and antiperiodic boundary conditions rather than the open boundary
conditions that must usually be used in real space.
Finally, using the k-DMRG the entanglement patterns of the wave
function can be analyzed with respect to
momentum, allowing us to make connections
between weak-coupling analytic pictures~\cite{Solyom} and numerical results.
In particular, we will be able to study the role of umklapp in one
dimension 
and the  effect of perfect and non-perfect nesting in two dimensions.

The paper is organized as follows.
In Sec.~\ref{sec:theory}, we describe
the model, define the quantum-information-based quantities that we
treat, and describe the numerical methods.
In Sec.\ \ref{sec:results}, we apply our approach to the
one- and two-dimensional Hubbard models in momentum space.
Finally, Sec.~\ref{sec:conclusion} contains our conclusions.

\section{Model and Methods}
\label{sec:theory}

\subsection{Hubbard model}
\label{sec:model}

The Hamiltonian of the Hubbard model in real space has the form
\begin{equation}
\label{eq:ham}
H = -t \sum_{\langle \boldsymbol{r},\boldsymbol{r}' \rangle,\sigma} 
\left( c_{\boldsymbol{r}\sigma}^{\dagger} 
c_{\boldsymbol{r}'\sigma}^{\phantom{\dagger}}
 + \rm{H.c.} \right) +
U \sum_{\boldsymbol{r}} n_{\boldsymbol{r}\uparrow} n_{\boldsymbol{r}\downarrow}\,,
\end{equation}
where $c_{\boldsymbol{r}\sigma}^{\dagger}$ 
and $c_{\boldsymbol{r}\sigma}^{\phantom{\dagger}}$
creates and annihilates a particle on site $\boldsymbol{r}$ 
with spin $\sigma=\{\downarrow,\uparrow\}$, 
$n_{\boldsymbol{r}\sigma}=c_{\boldsymbol{r}\sigma}^{\dagger}
c_{\boldsymbol{r}\sigma}^{\phantom{\dagger}}$ 
measures the number of electrons on site $\boldsymbol{r}$ 
with spin $\sigma$, 
and $\langle \boldsymbol{r}, \boldsymbol{r}' \rangle$ denotes
a nearest-neighbor pair on sites $\boldsymbol{r}$ and $\boldsymbol{r}'$.
The dimensionality and structure of the lattice is contained in the
the definition of the index $\boldsymbol{r}$ 
and in the
definition of the nearest neighbors.
By carrying out a change of basis using a Fourier transformation,
i.e., $c_{\boldsymbol{k}\sigma}^{\phantom{\dagger}}=1/\sqrt{N}\sum_{\boldsymbol{r}}
\exp(\text{i} \, \boldsymbol{k} \cdot \boldsymbol{r})\,
 c_{\boldsymbol{r}\sigma}^{\phantom{\dagger}}$, we obtain the corresponding
momentum-space Hamiltonian
\begin{equation}
\label{eq:ham-k}
H = \sum_{\boldsymbol{k}\sigma} \varepsilon(\boldsymbol{k}) 
c_{\boldsymbol{k}\sigma}^{\dagger} 
c_{\boldsymbol{k}\sigma}^{\phantom{\dagger}}+ 
\frac{U}{N} \sum_{\boldsymbol{k}\boldsymbol{p}\boldsymbol{q}} 
c_{\boldsymbol{p}-\boldsymbol{q}\uparrow}^{\dagger} c_{\boldsymbol{k}+\boldsymbol{q}\downarrow}^{\dagger} 
c_{\boldsymbol{k}\downarrow}^{\phantom{\dagger}}
c_{\boldsymbol{p}\uparrow}^{\phantom{\dagger}}\,,
\end{equation}
if periodic boundary conditions are applied,
where $c_{\boldsymbol{k}\sigma}^{\dagger}$ and
$c_{\boldsymbol{k}\sigma}^{\phantom{\dagger}}$ creates and annihilates a particle with
momentum $\boldsymbol{k}$, with $\boldsymbol{k}$
a vector with the dimensionality of the lattice, and $N$ is the number of sites.
In one dimension, $\boldsymbol{k} = k = (2\pi n)/N$, with $-N/2< n \leq N/2$ and 
$\varepsilon(k)=-2t\cos(k)$.
In two dimensions, ${\boldsymbol{k}}=(k_x,k_y)$ and
$\varepsilon(\boldsymbol{k})=-2t\cos(k_x)-2t\cos(k_y)$. 
Note that the $\boldsymbol{k}$~points will be shifted by a half-interval if
antiperiodic rather than periodic boundary conditions 
are applied in a particular direction.
Due to momentum conservation, the total momentum before and after
the scattering processes given by the four-operator term is the same. 
The interaction strength $U$ is given in units of $t$ with $t=1$.

In the remainder of the paper, we will use the expression ``Hartree-Fock
orbitals''  to denote single-particle orbitals, i.e., $\boldsymbol{k}$~points, that
are filled for the noninteracting case,  $U=0$.
These points will be within the Fermi surface 
where the ``Fermi surface'' is defined
by the boundary in momentum space between occupied and unoccupied
points for $U=0$.
In one dimension, the Fermi surface consists of two points, 
$\pm k_F$, and in two dimensions, it forms a 
curve or set of curves in
the $k_x$-$k_y$ plane in the infinite-system limit.
Also note that we measure distance in momentum space 
in units of the unit vectors of the reciprocal space,
that is, $2\pi/N$ for the one-dimensional chain 
and $2\pi/L_x$ or $2\pi/L_y$ for the two-dimensional
square lattice of dimension $L_x\times L_y = N$. 
Use of terms such as ``long range'' or ``short range'' as applied to
momentum space will refer to momentum-space distances measured in these units.
``Volume'' in momentum space will be shorthand for the number of discrete
momentum points in a momentum region or in the entire lattice.

\subsection{DMRG and entanglement-based measures}
\label{sec:dmrg}

In the two-site DMRG, the full 
Hilbert space of a finite
system consisting of $N$ sites, $\Lambda^{(N)}=\otimes_{i=1}^N \Lambda_i$,
is approximated by a tensor product of four tensor spaces,
$\Xi^{(N)}_{\rm DMRG}
=\Xi^{(\mathrm{l})}\otimes\Lambda_{l+1}\otimes\Lambda_{l+2}
\otimes\Xi^{(\mathrm{r})}$. 
The basis states of $\Lambda_i$, the local-site (tensor) space,
depend on the representation in which the Hamiltonian is formulated.
The bases of the $\Xi^{(\mathrm{l})}$ and  $\Xi^{(\mathrm{r})}$
are formed through a series of unitary transformations generated by
singular value decompositions (SVD) 
applied repeatedly as the bipartite partition of the system is moved
back and forth through the lattice (``sweeping").
Thus, their 
actual form depends on the
details of the procedure~\cite{Schollwock-2005,Noack-2005,Hallberg-2006,Schollwock-2011,Szalay-2015}.
The nature and size of the bases of $\Xi^{(\mathrm{l})}$ and $\Xi^{(\mathrm{r})}$ 
can be optimized using the concepts of quantum information 
theory~\cite{Nielsen-2000,Preskill,Wilde-2013}
i.e., by controlling the level of 
entanglement~\cite{Horodecki-2009,Szalay-2013,Eisert-2010} of the
subsystems~\cite{Vidal-2003,Legeza-2003a,Legeza-2003c,Rissler-2006,PEPS,Szalay-2015}.
Therefore, a given problem can
be solved with significantly less computational resources if an 
optimal representation is 
found~\cite{Rissler-2006,Murg-2010-tree,Fertitta-2014,Krumnow-2015}.

The single-site von Neumann entropy,
$s_i=-{\rm Tr}\rho_i \ln \rho_i$ is formed from the reduced density 
matrix $\rho_i$ of the subsystem consisting of the site $i$. 
Its value ranges from $0$ to $\ln 4$ for a fermionic site with
spin, with larger values corresponding to a more mixed state and a
larger contribution to the correlation energy.
In fact, an exact expression for the local single-site entropy in the 
ground state of the one-dimensional Hubbard model has also been
derived~\cite{Larsson-2005}.
Similarly, the two-site von Neumann entropy can be constructed
using the reduced density matrix of a subsystem
built from orbitals $i$ and $j$, $\rho_{ij}$.
The mutual information
$I_{ij}=s_i+s_j-s_{ij}$ describes how these two orbitals are correlated 
with each other given that they are embedded in the whole system.
It includes contributions from correlations of
both classical and quantum nature~\cite{Modi-2010}.
The number of block states, $M_l={\rm \dim}\ \Xi^{(\mathrm{l})}$
and $M_r={\rm \dim}\ \Xi^{(\mathrm{r})}$, required to achieve
a given convergence is determined by the 
von Neumann entropy of segments of length $l=1\ldots N-1$
of the finite chain (``blocks'')~\cite{Vidal-2003,Legeza-2003a}.
In practice, we fix the accuracy threshold and vary the number of block states
$M_{\rm max} = \max{(M_l,M_r)}$\ to maintain the threshold.
The $l$-site block entropy $S(l)$ can be used to study critical and
gapped phases and is an easily accessed quantity numerically, as the 
reduced density matrices $\rho_{l}$ are generated automatically within
the DMRG procedure. 

Ground states of lattice models with local interactions are thought to
obey the so-called entropy area 
law relatively generically~\cite{Eisert-2010}.
In one dimension, the block entropy can be shown to 
saturate with block length for gapped models and to diverge logarithmically
for critical systems~\cite{Vidal-2003}.
In two dimensions, the entropy area law still applies 
if interactions are short-range and all 
correlation functions have finite correlation lengths~\cite{Wolf-2008}.
When nonlocal interactions are present, the arguments for the
applicability of the area law break down. 
In nonlocal models, the profile of the block
entropy has a more general form
that depends on several factors, as has been discussed
for the Hubbard model in momentum space~\cite{Legeza-2003c},
for the XXZ chain in momentum space~\cite{Lundgren-2014},
and for quantum chemical systems~\cite{Barcza-2011}.
The block entropy profile can be optimized by minimizing the entanglement
distance, defined as
\begin{equation}
I_{\rm dist} = \sum_{ij} I_{ij}d^{\eta}\,,
\label{eq:Idist}
\end{equation}
where the sum is weighted by the $\eta^{\rm th}$ moment of the
distance ($\eta\ge 1$)~\cite{Rissler-2006,Barcza-2011,Fertitta-2014},
and the distance function $d$
depends on the tensor network topology~\cite{Murg-2015-tree}.
In the MPS--based DMRG, 
$d=|i-j|$ refers to the distance between sites within the MPS chain,
 and Eq.~(\ref{eq:Idist}) can be minimized
by permuting the lattice sites, i.e., by 
reordering~\cite{Legeza-2003c,Rissler-2006,Barcza-2011}.
In this way, both the sum of $S(l)$ over $l=1\ldots N-1$ and the 
maximum of $S(l)$ over all $l$, $\max[S(l)]$, can be reduced significantly. 
The sum is related to the computational time of a full sweep, and the
maximum determines the maximum amount of computational resources required for 
one DMRG step.
By changing a local hopping to a nonlocal hopping by permuting the
ordering of lattice sites, one can interpolate between a logarithmic-
(area law) and a linear (volume law) 
scaling~\cite{Gori-2015}, as has been demonstrated
in a study of the one-dimensional Hubbard model~\cite{Legeza-2006}.

Another way of reducing $I_{\rm dist}$ is to transform the
single-particle basis, choosing the transformation to optimize
entanglement measures~\cite{Legeza-2003c,Murg-2010-tree,Krumnow-2015}.
For example, for the noninteracting ($U=0$) Hubbard model, 
the Hamiltonian is diagonal in the momentum-space representation,
$s_{i}=0$, $I_{i,j}=0$, and the ground state is a
product state; thus, $I_{\rm dist}$ and $S(l)$ are zero.
On the other hand, for finite Coulomb interaction $U$, 
these quantities become finite and, in the $U\rightarrow\infty$ limit,
$s_{i}=\ln 4$, i.e., all sites are in maximally mixed states.
Therefore, the momentum-space representation is expected to be useful
in a regime adiabatically connected to the free-particle case,
typically the weak-coupling regime.

Entanglement analysis is also a very powerful tool to obtain physical
information encoded in the wave function~\cite{Boguslawski-2012b,
Boguslawski-2013a,Kurashige-2013,Barcza-2014,Fertitta-2014}.
Entanglement patterns  by the two-site mutual information
provide information about the overall correlations in the system,
while the generalized correlation functions used to form the two-site
reduced density matrix can be used to identify 
dominant correlations.
For more detailed discussions, 
see Refs.~\onlinecite{Rissler-2006,Barcza-2014}.

\subsection{Momentum-space DMRG implementation}

Some aspects of the usual DMRG algorithm must be adapted to obtain an
efficient implementation for momentum-space Hamiltonians.

The main difficulty in implementing the momentum-space DMRG comes from
the fact that short-range interaction terms such as that in the
real-space Hubbard model, Eq.~\eqref{eq:ham},
become long-range in the momentum basis.
Naively implemented, the momentum-space Hubbard model Hamiltonian
requires the calculation of $\mathcal{O}\left(N^3\right)$ terms in
order to apply the interaction part of the Hamiltonian to the wave
vector because
of sum over three momenta in Eq.~\eqref{eq:ham-k}.
In an MPS formulation, the matrix product operator (MPO) would have 
bond dimension $M \propto N^3$, 
resulting in unacceptable resource requirements even for relatively
small system sizes. 
This problem can be ameliorated by factorizing the Hamiltonian and
building partial sums over certain
combinations of operators within the left and right DMRG blocks,
resulting in an effective one-index sum and a computational cost per
diagonalization step of $\mathcal{O}\left(N\right)$.
At the same time, the required composed operators must be saved and
updated iteratively,
resulting in additional memory cost of $\mathcal{O}\left(N\right)$.
While this technique was already described and used in the first work on
momentum-space DMRG~\cite{Xiang-1996},
here we have adapted this method to the MPS
framework and optimized it within this framework.

The factorization described above can only be carried out for
particular forms of the interaction of which the Fourier-transformed
on-site interaction is one.
The CPU time and memory costs for a single diagonalization
step can be reduced to $\mathcal{O}\left(N\right)$,
compared to $\mathcal{O}\left(N^2\right)$ for other more general
long-range Hamiltonians (e.g., those for quantum chemistry)~\cite{White-1999}.
Certain extensions to the Hubbard interaction, such as nearest- or
next-nearest-neighbor repulsion can also be factorized to
obtain $\mathcal{O}\left(N\right)$ terms, 
only affecting the prefactor but not the scaling in the computational
cost~\cite{Xiang-1996}.
In contrast to the real-space case, adding longer-range terms to
the hopping in momentum space only 
affects the dispersion relation, which enters the Hamiltonian as a
diagonal term, and therefore does not change
the MPO bond dimension.

One significant advantage of working in the momentum-space
representation is the availability of
the additional momentum quantum numbers.
Within an MPS--based DMRG implementation, all contractions of the
tensor network can be reduced to a series of matrix operations,
principally multiplications.
The Abelian momentum quantum numbers can be used to
decompose all matrices to block form, where the blocks are labeled by
quantum number pairs.
Carrying out operations only using these blocks significantly speeds up
the algorithm and significantly lowers the memory required.
For the Hubbard model calculations carried out here, we typically
obtain very small dense blocks
with $\mathcal{O}\left(10{\times}10\right)$ elements, 
which make up sparse matrices with up to millions of blocks.
This large number of blocks makes it essential to carry out the
quantum-number bookkeeping as efficiently as possible and to reduce
additional overhead to a minimum.
In our code, we obtain very good results by using hash tables to do the 
quantum-number bookkeeping.

At the same time, use of the momentum quantum numbers introduces
convergence problems because the quantum numbers of local states of
two neighboring sites can, in general, no longer be recombined to the
same target quantum number in more than one way.
Due to this, the algorithm tends to get stuck in local minima even if the DMRG
is implemented in the usual way with two single sites in the center.
(Such problems typically occur in the variant of the DMRG where a
single site is taken in the center.)
For example, a two-site k-DMRG calculation initialized with a product
state will remain stuck in that state throughout the entire
calculation unless additional measures are taken.
For the single-site real-space DMRG algorithm, an effective solution
to this problem is to add a noise term to the density
matrix~\cite{White-2005}.
Here, we add such a noise term within the two-site algorithm and find
that the convergence problem is essentially eliminated, provided that
the noise term is given appropriate strength.

In our implementation, which exploits the additional momentum
quantum numbers and carries out an efficient factorization of the
interaction term, a comparatively large
number of block states can be kept at reasonable computational cost.
We typically increase the number of block states
after every second full sweep by $4\,000$ states 
and keep a maximum of up to $54\,000$ states
during the last two sweeps.
Running on 10 CPU cores (Intel\textsuperscript{\textregistered}
Xeon\textsuperscript{\textregistered} 5140), 
the calculation for a $6{\times}6$-site system 
with a maximum of $54\,000$ states kept during the last two sweeps
takes roughly three days of wall-clock time 
and uses 30 GB of main memory.
We use two codes, one sequential code based on a 
strongly modified version of the open-source ITensor library~\cite{itensor}
and one parallelized code based on a self-written tensor library.

In addition to these technical optimizations, 
choosing appropriate boundary conditions for each calculation
so that a ``closed-shell'' configuration is obtained for the
Fermi sea
can lower the maximum block entropy and thus
improve the DMRG convergence for a fixed number of block states.
The idea is to coordinate lattice size, filling, and boundary conditions
so that the noninteracting ground state is non degenerate.
For example, consider the  
noninteracting half-filled 
one-dimensional Hubbard model in momentum space.
For periodic boundary conditions
and system sizes $N \in \left\{ 2,6,10,14,..\right\}$
the ground state in momentum space is a unique product state,
while system sizes $N \in \left\{ 4,8,12,16,..\right\}$
lead to a fourfold-degenerate ground state.
For antiperiodic boundary conditions, the momentum-space sites are shifted, 
reversing the situation.
It is known that this behavior is preserved in the exact Bethe
ansatz solution for the interacting one-dimensional system~\cite{essler_book}.
Although interaction tends to wash out degeneracies in open-shell
configurations, taking closed-shell configurations for weak to
moderate coupling strength is still very useful.
For moderate coupling strength, closed-shell configurations
have a unique Hartree-Fock ground state and
thus have fewer energetically low-lying excitations,
leading to a less entangled ground state.
The system sizes, fillings, and boundary conditions in 
the next section are selected accordingly.

\section{Results}
\label{sec:results}

In this section, we present our numerical results obtained with the
momentum-space DMRG on the half-filled and doped Hubbard model
in one and two dimensions for finite systems.
First, we analyze the entanglement patterns in the ground-state wave
function, identifying relevant correlations by examining
highly entangled momenta points. 
Next, we calculate the wave-function coefficients in full tensor form,
which are an alternate way of characterizing the wave function and can
be connected to the configuration-interaction (CI) expansion
technique used in quantum chemistry.
Finally, we study the scaling of entanglement,
quantified by the von Neumann block entropy, as a function 
of interaction strength and system size. 

\subsection{Entanglement patterns and correlations}
\label{sec:entanglement_patterns}

In order to investigate the entanglement structure 
and optimize the site ordering within the MPS, 
we examine the two-site mutual information $I_{ij}$ 
as well as the site occupancy and single-site entropy 
for the one- and two-dimensional Hubbard models.

\begin{figure*}
\centering
\includegraphics[width=18.5cm]{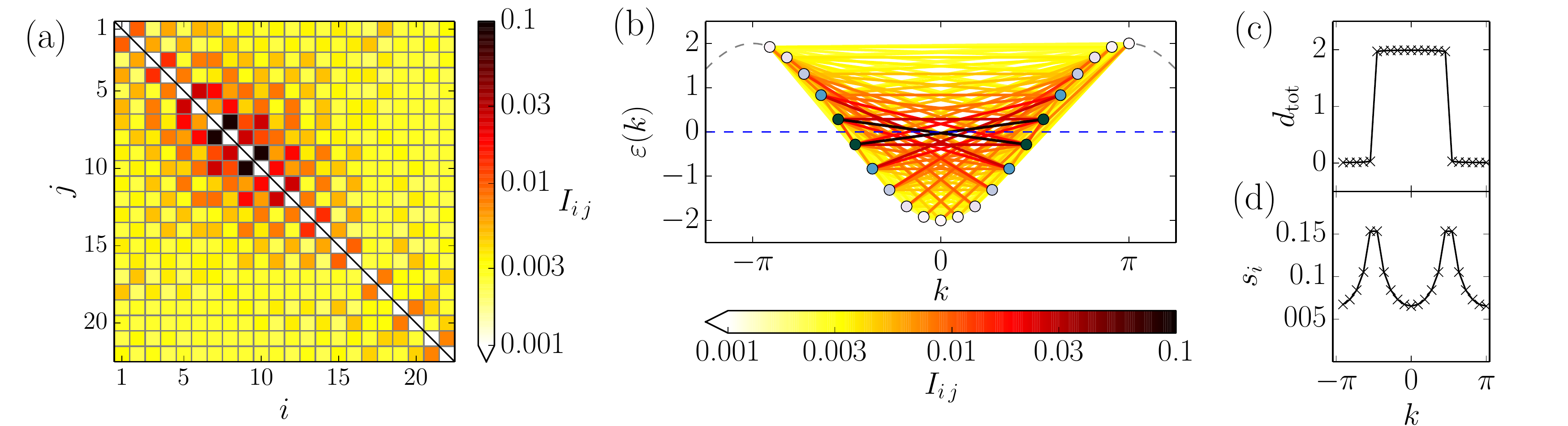}
\caption{
(Color online) 
Ground state of the $N=22$-site Hubbard model in momentum space 
(with periodic boundary conditions) 
for $U=1.0$ at half-filling.
(a) Two-site mutual information $I_{i,j}$ between MPS sites $i$ and $j$ 
for the optimal site ordering (Table~\ref{tab:1d_ordering}).
(b) Two-site mutual information $I_{i,j}$ between momentum points $k$, where
the $k$~points are arranged according to the dispersion relation. 
The blue (gray) dashed line indicates the Fermi level. 
(c) Site occupancy $d_{\mbox{tot}}$. 
(d) Single-site entropy $s_{i}$.
}
\label{fig:ground_state_1d_n1}
\end{figure*}

\begin{figure}
\includegraphics[width=8.0cm]{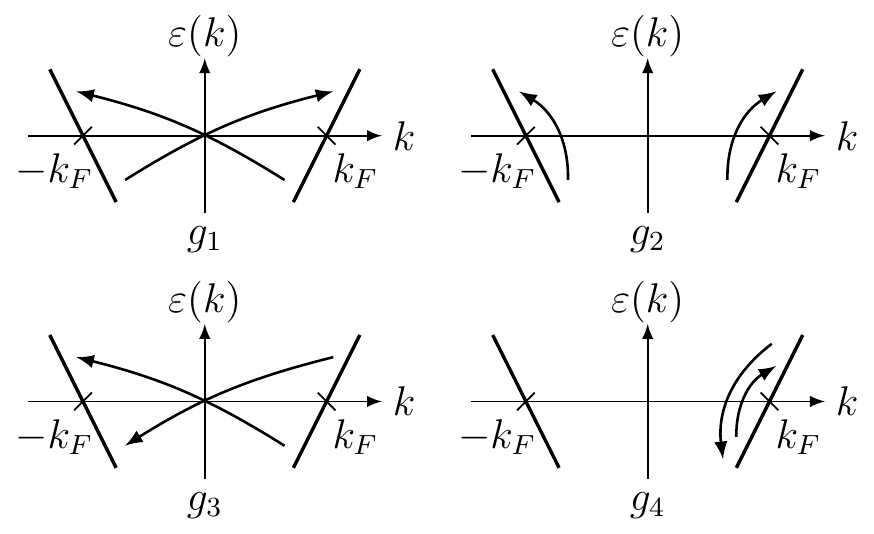}
\caption{
Scattering processes near the Fermi surface
in the half-filled system where
$g_1$ is backward scattering,
$g_2$ and $g_4$ forward scattering, and
$g_3$ umklapp scattering.
}
\label{fig:g-ology}
\end{figure}

We start with the half-filled one-dimensional case in
Fig.~\ref{fig:ground_state_1d_n1}.
Figure~\ref{fig:ground_state_1d_n1}(a) is a (color) density plot of the
$I_{ij}$ matrix, with the 
row and column indices $i$ and $j$ referring to the sites of the DMRG chain, 
while in Fig.~\ref{fig:ground_state_1d_n1}(b) the same correlations are 
plotted as lines between the corresponding points in momentum space,
where the momentum points are arranged in
the vertical direction according to the  
dispersion relation $\varepsilon(k)$.
One can see that the pairs of sites with the strongest correlations
are the pairs which are associated with the allowed scattering processes
close to the Fermi surface, as depicted in Fig.~\ref{fig:g-ology}.
Consequently, the single-site entropy $s_i$
[Fig.~\ref{fig:ground_state_1d_n1}(d)] shows two peaks at the Fermi
points. Correspondingly, the site occupancy
[Fig.~\ref{fig:ground_state_1d_n1}(c)] drops from
close to $2.0$ to almost $0.0$ at the Fermi points; the sharp drop is
characteristic of the relatively weak interaction strength $U=1.0$. 
The strongest bonds can be found between those pairs of sites 
with distance $k_1-k_2=\pi$ in momentum space.
Therefore, in order to improve the DMRG convergence,
sites with a separation $\pi$ in momentum space should be placed on 
neighboring sites on the DMRG chain.
In addition, pairs close to the Fermi surface should be put in the center
of the DMRG chain. 
Automatic ordering of the sites, carried out by
minimizing the entanglement distance (\ref{eq:Idist}), gives the
optimal ordering shown in Table~\ref{tab:1d_ordering},
which corresponds quite well to these heuristic rules.

\begin{figure*}
\centering
\includegraphics[width=18.5cm]{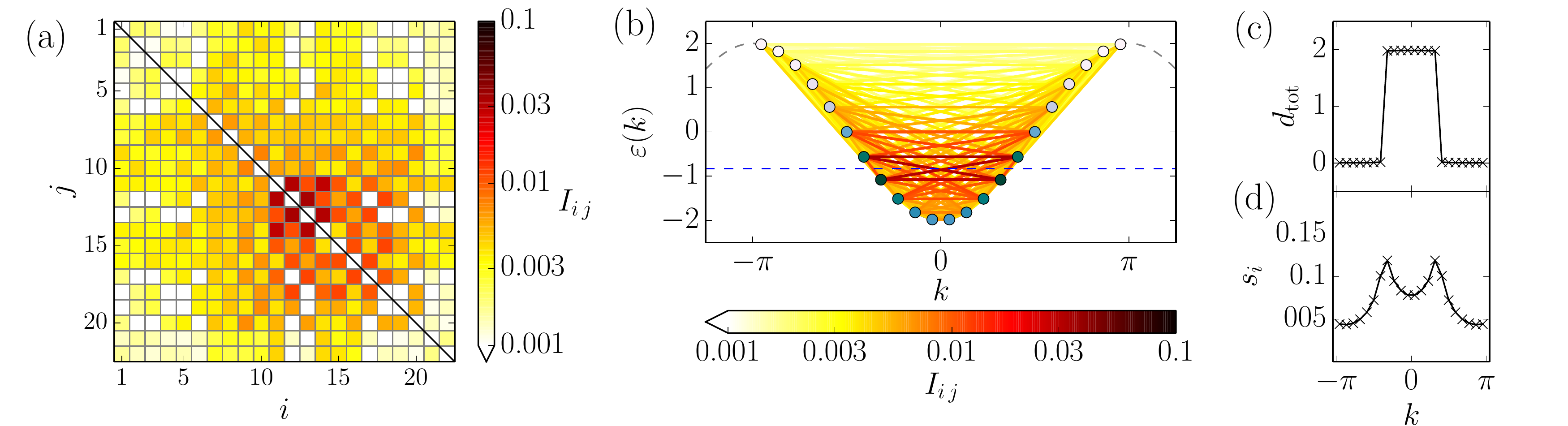}
\caption{
(Color online) 
Ground state of the doped $N=22$-site Hubbard model in momentum space 
(with antiperiodic boundary conditions) 
for $U=1.0$ at average occupancy $n=0.727$.
(a) Two-site mutual information $I_{i,j}$ between MPS sites $i$ and $j$ 
for the optimal site ordering (Table~\ref{tab:1d_ordering}).
(b) Two-site mutual information $I_{i,j}$ between momentum points $k$,
where the $k$~points are arranged according to the dispersion relation. 
The (blue) dashed line indicates the Fermi level. 
(c) Site occupancy $d_{\mbox{tot}}$. 
(d) Single-site entropy $s_i$.
}
\label{fig:ground_state_1d_n075}
\end{figure*}

We now examine the doped one-dimensional system
[Fig.~\ref{fig:ground_state_1d_n075}], taking an average
particle number $n=0.727$.
The entanglement structure is similar to that in the half-filled case;
the pairs of sites close to and symmetric relative to the  Fermi points
(now at somewhat smaller $k$) show the strongest correlations [see
Fig.~\ref{fig:ground_state_1d_n075}(b)].
However, the pairs of momentum points analogous to those which, in the
half-filled case, are associated with backward and umklapp scattering,
$g_1$ and $g_3$, respectively, are now significantly less correlated.
This can be understood by examining the effect of the shifting of the
Fermi points  on the conservation of momentum.
In the half-filled case,
the energetically lowest excitations relative to the 
Hartree-Fock ground state $\left| \Psi \right\rangle_{\rm{HF}}$ that have the same total
momentum are of four different types:
\begin{subequations}
\begin{align}
 \left|\Psi\right\rangle_A & = 
 c^{\dag}_{\pm k_2  \downarrow} 
 c^{\dag}_{\pm k_2  \uparrow} 
 c_{ \mp k_1  \uparrow} 
 c_{ \mp k_1  \downarrow} 
 \left| \Psi \right\rangle_{\rm{HF}} \, ,
  \\
 \left|\Psi\right\rangle_B & = 
 c^{\dag}_{k_2  \sigma} 
 c^{\dag}_{-k_2  \bar{\sigma}} 
 c_{k_1  \sigma} 
 c_{-k_1  \bar{\sigma}} 
 \left| \Psi \right\rangle_{\rm{HF}} \, ,
  \\
 \left|\Psi\right\rangle_C & = 
 c^{\dag}_{k_2  \sigma} 
 c^{\dag}_{-k_2  \bar{\sigma}} 
 c_{k_1  \bar{\sigma}} 
 c_{-k_1  \sigma} 
 \left| \Psi \right\rangle_{\rm{HF}} \, ,
 \end{align}
 \text{and} 
 \begin{align}
  \left|\Psi\right\rangle_D  = 
  c^{\dag}_{k_2  \sigma} 
  c^{\dag}_{-k_2  \sigma} 
  c_{k_1  \sigma} 
  c_{-k_1  \sigma} 
  \left| \Psi \right\rangle_{\rm{HF}} \, ,
   \end{align}
\end{subequations}
with $k_1 = k_F-\delta_k$, $k_2 = k_F+\delta_k$, $\delta k = \pi/N$,
and $\sigma \in \left\{ \uparrow , \downarrow \right\}$.  
In the doped case, states of type $A$ no longer have the same momentum
as $\left|\Psi\right>_{\rm{HF}}$ because $k_F\neq\pi/2$.
Thus, they cannot be in the space of the ground state.
Without these excitations, two-particle umklapp processes symmetric to
the Fermi surface are no longer possible; this 
also holds for higher excitations.
In terms of equal-time correlations, the number of excitations
that make up the interacting ground state
and can contribute to the correlations drops, consequently lowering
the two-site mutual information between these sites.

In matrix form for optimal ordering, $I_{ij}$
[Fig.~\ref{fig:ground_state_1d_n075}(a)] is somewhat less
diagonal-dominated than at half-filling, but the strongly off-diagonal
elements have comparable or smaller values than the half-filled case.
Note that although the Fermi edge is quite sharp
[Fig.~\ref{fig:ground_state_1d_n075}(c)] the peaks in the single-site
entropy $s_i$ [Fig.~\ref{fig:ground_state_1d_n075}(d)] are somewhat
washed out relative to the half-filled case, reflecting the reduced
entanglement of momenta near the Fermi points.

\begin{table}
\small
\centering
\begin{tabular}{|
>{\centering\arraybackslash}p{0.75cm} || 
>{\centering\arraybackslash}p{0.75cm} | 
>{\centering\arraybackslash}p{1.25cm} || 
>{\centering\arraybackslash}p{0.75cm} | 
>{\centering\arraybackslash}p{1.25cm} |}
\hline
 & \multicolumn{2}{c||}{n=1.0} & \multicolumn{2}{c|}{n=0.727} \\
\hline
$i$ & $\text{k}$ & $\epsilon(k)$ & $\text{k}$ & $\epsilon(k)$ \\
\hline
1   & 9   &  1.683 & 10  &  1.819 \\
2   & 20  & -1.683 & 13  &  1.819 \\
3   & 8   &  1.31  & 14  &  1.511 \\
4   & 19  & -1.31  & 9   &  1.511 \\
5   & 7   &  0.831 & 8   &  1.081 \\
6   & 18  & -0.831 & 15  &  1.081 \\
7   & 17  & -0.285 & 1   & -1.98  \\
8   & 6   &  0.285 & 0   & -1.98  \\
9   & 16  &  0.285 & 2   & -1.819 \\
10  & 5   & -0.285 & 21  & -1.819 \\
11  & 15  &  0.831 & 4   & -1.081 \\
12  & 4   & -0.831 & 18  & -0.564 \\
13  & 14  &  1.31  & 5   & -0.564 \\
14  & 3   & -1.31  & 19  & -1.081 \\
15  & 13  &  1.683 & 20  & -1.511 \\
16  & 2   & -1.683 & 3   & -1.511 \\
17  & 10  &  1.919 & 17  &  0.0   \\
18  & 21  & -1.919 & 6   &  0.0   \\
19  & 1   & -1.919 & 7   &  0.564 \\
20  & 12  &  1.919 & 16  &  0.564 \\
21  & 11  &  2.0   & 11  &  1.98  \\
22  & 0   & -2.0   & 12  &  1.98  \\
\hline
\end{tabular}
\caption{ 
Mapping between MPS-site indices~$i$
and momentum-space site indices~$\text{k}$  
for the optimal ordering
of the of the $22$-site system
depicted in Figs.~\ref{fig:ground_state_1d_n1}
and~\ref{fig:ground_state_1d_n075},
with momentum points 
$ k = \text{k} \frac{2 \pi }{22} $
for half-filling $n=1.0$ and shifted momentum points
$ k = \left( \text{k} + 0.5 \right) \frac{2 \pi }{22} $
for the doped case $n=0.727$.
}
\label{tab:1d_ordering}
\end{table}

\begin{figure*}
\centering
\includegraphics[width=18.0cm]{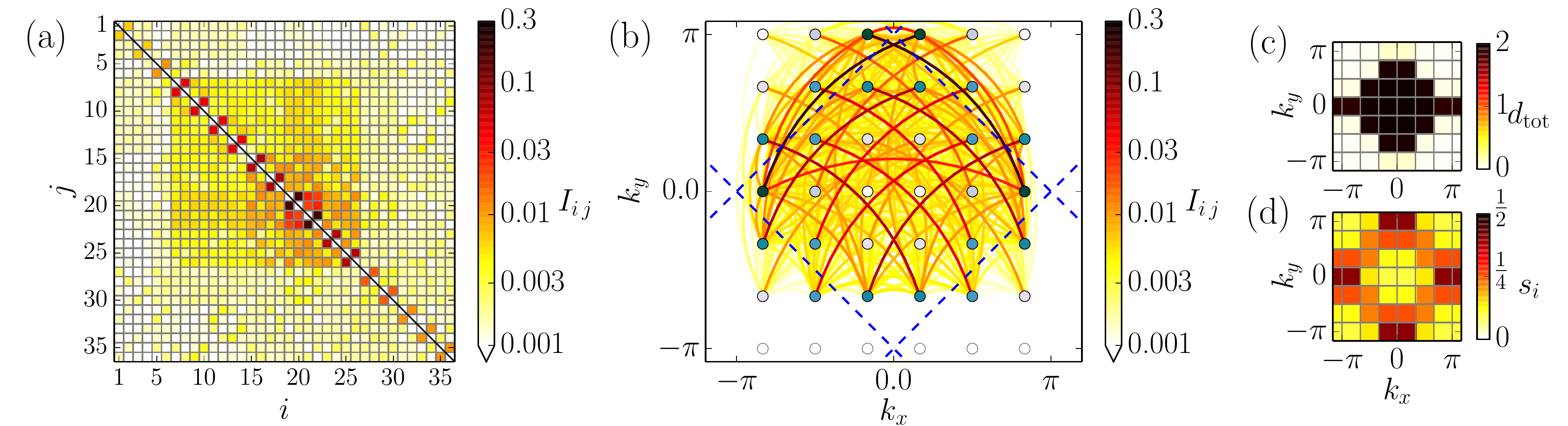}
\caption{
(Color online)
Ground state of the $6{\times}6$ Hubbard model in momentum space
(antiperiodic / periodic boundary conditions in $x$- / $y$-direction) 
for $U=2.0$ at half-filling: 
(a) two-site mutual information $I_{i,j}$ between MPS sites $i$ and $j$ 
for the optimal site ordering (Table~\ref{tab:2d_ordering}).
(b) two-site mutual information $I_{i,j}$ between momentum points $(k_x,k_y)$.
The (blue) dashed line indicates the Fermi surface. 
(c) Site occupancy $d_{\mbox{tot}}$. 
(d) Single-site entropy $s_i$.
}
\label{fig:ground_state_2d_n1}
\end{figure*}

\begin{figure*}
\centering
\includegraphics[width=18.0cm]{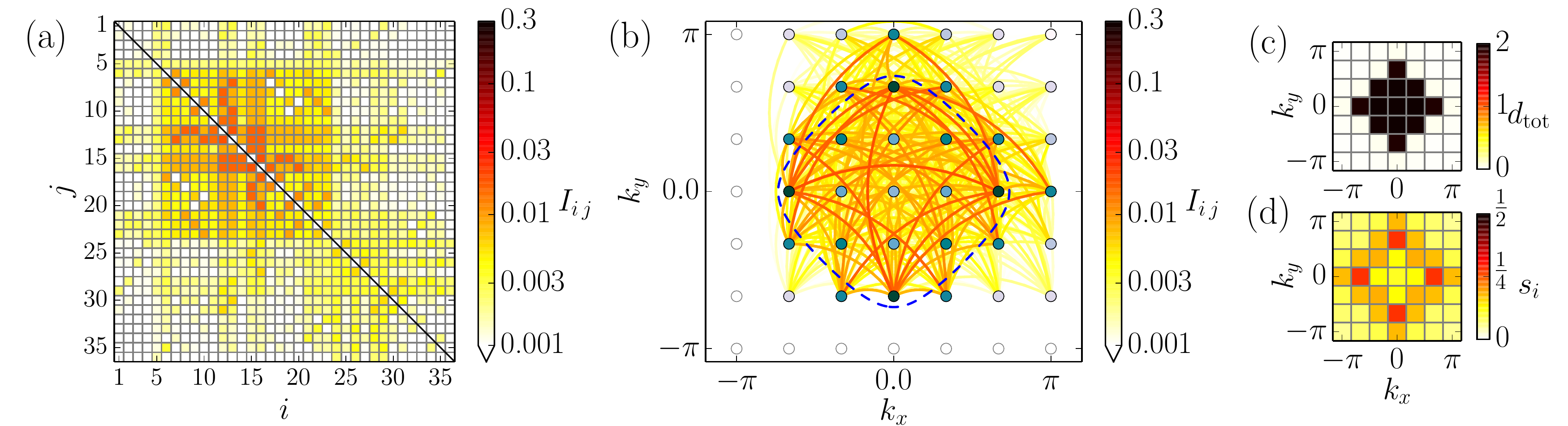}
\caption{
(Color online)
Ground state of the doped $6{\times}6$ Hubbard model in momentum space
(periodic boundary conditions in both directions) 
for $U=2.0$ at $n=0.722$ filling: 
(a) two-site mutual information $I_{i,j}$ between MPS sites $i$ and $j$ 
for the optimal site ordering (Table~\ref{tab:2d_ordering}).
(b) two-site mutual information $I_{i,j}$ between momentum points $(k_x,k_y)$.
The (blue) dashed curve indicates the Fermi surface. 
(c) Site occupancy $d_{\mbox{tot}}$. 
(d) Single-site entropy $s_i$.
}
\label{fig:ground_state_2d_n075}
\end{figure*}

\begin{table}
\small
\centering
\begin{tabular}{|
>{\centering\arraybackslash}p{0.75cm} || 
>{\centering\arraybackslash}p{0.5cm} | 
>{\centering\arraybackslash}p{0.5cm} | 
>{\centering\arraybackslash}p{1.25cm} || 
>{\centering\arraybackslash}p{0.5cm} | 
>{\centering\arraybackslash}p{0.5cm} | 
>{\centering\arraybackslash}p{1.25cm} |}
\hline
 & \multicolumn{3}{c||}{n=1.0} & \multicolumn{3}{c|}{n=0.722} \\
\hline
$i$ & $ \text{k}_x$ & $ \text{k}_y$ & $\epsilon(\boldsymbol{k})$ 
& $ \text{k}_x$ & $ \text{k}_y$ & $\epsilon(\boldsymbol{k})$ \\
\hline
1   & 2  & 3  &  3.732 & 3  & 2   &  3.0 \\
2   & 5  & 0  & -3.732 & 3  & 4   &  3.0 \\
3   & 3  & 3  &  3.732 & 2  & 3   &  3.0 \\
4   & 0  & 0  & -3.732 & 4  & 3   &  3.0 \\
5   & 2  & 2  &  2.732 & 0  & 0   & -4.0 \\ 
6   & 5  & 5  & -2.732 & 5  & 5   & -2.0 \\
7   & 1  & 2  &  1.0   & 2  & 1   &  0.0 \\
8   & 4  & 5  & -1.0   & 4  & 5   &  0.0 \\
9   & 4  & 4  &  1.0   & 2  & 5   &  0.0 \\ 
10  & 1  & 1  & -1.0   & 4  & 1   &  0.0 \\ \hline
11  & 4  & 2  &  1.0   & 3  & 0   &  0.0 \\ 
12  & 1  & 5  & -1.0   & 0  & 2   & -1.0 \\ 
13  & 1  & 4  &  1.0   & 0  & 4   & -1.0 \\ 
14  & 4  & 1  & -1.0   & 0  & 3   &  0.0 \\ 
15  & 3  & 1  &  0.732 & 2  & 0   & -1.0 \\ 
16  & 0  & 4  & -0.732 & 4  & 0   & -1.0 \\ 
17  & 5  & 4  & -0.732 & 1  & 2   &  0.0 \\ 
18  & 2  & 1  &  0.732 & 5  & 4   &  0.0 \\ 
19  & 2  & 0  & -0.268 & 5  & 2   &  0.0 \\ 
20  & 5  & 3  &  0.268 & 1  & 4   &  0.0 \\ \hline
21  & 3  & 0  & -0.268 & 1  & 1   & -2.0 \\ 
22  & 0  & 3  &  0.268 & 5  & 1   & -2.0 \\ 
23  & 2  & 5  &  0.732 & 1  & 5   & -2.0 \\ 
24  & 5  & 2  & -0.732 & 1  & 0   & -3.0 \\ 
25  & 3  & 5  &  0.732 & 5  & 3   &  1.0 \\ 
26  & 0  & 2  & -0.732 & 5  & 0   & -3.0 \\ 
27  & 4  & 0  & -2.0   & 1  & 3   &  1.0 \\ 
28  & 1  & 3  &  2.0   & 0  & 1   & -3.0 \\ 
29  & 4  & 3  &  2.0   & 0  & 5   & -3.0 \\ 
30  & 1  & 0  & -2.0   & 3  & 1   &  1.0 \\ \hline
31  & 3  & 2  &  2.732 & 3  & 5   &  1.0 \\ 
32  & 0  & 5  & -2.732 & 4  & 2   &  2.0 \\ 
33  & 2  & 4  &  2.732 & 2  & 4   &  2.0 \\ 
34  & 5  & 1  & -2.732 & 4  & 4   &  2.0 \\
35  & 3  & 4  &  2.732 & 2  & 2   &  2.0 \\ 
36  & 0  & 1  & -2.732 & 3  & 3   &  4.0 \\
\hline
\end{tabular}
\caption{ 
Mapping between and MPS-site indices~$i$
and  momentum-space site indices~$\left\{\,\text{k}_x\,,\,\text{k}_y\,\right\}$  
for the optimal ordering
of the $6{\times}6$ system 
depicted in 
Figs.~\ref{fig:ground_state_2d_n1}
and~\ref{fig:ground_state_2d_n075},
with momentum points 
$ \left(\,k_x\,,\,k_y\,\right) = \left(\,\left( \text{k}_x + 0.5 \right)\frac{2 \pi }{6}  \,,\, \text{k}_y\frac{2 \pi }{6}\,\right)$
for half-filling $n=1.0$ and 
$ \left(\,k_x\,,\,k_y\,\right) = \left(\,\text{k}_x \frac{2 \pi }{6} \,,\, \text{k}_y \frac{2 \pi }{6}\,\right)$
for the doped case, $n=0.722$.
}
\label{tab:2d_ordering}
\end{table}

We now proceed to the entanglement patterns 
in the two-dimensional Hubbard model.
Many characteristics observed in the one-dimensional model can be found
in generalized form in the two-dimensional case.
At half-filling, we consider points $\boldsymbol{k}_F$ and
$\tilde{\boldsymbol{k}}_F$ that lie on opposite sides of the Fermi
surface, i.e., that fulfill the nesting condition 
$\boldsymbol{k}_F=\tilde{\boldsymbol{k}}_F+\left(\pi,\pi\right)$.
Pairs of lattice sites directly above and below such opposing
locations on the Fermi surface also fulfill the nesting condition 
and allow for energetically favorable processes 
that are generalizations of umklapp processes in the one-dimensional
system.
As can be seen in Figs.~\ref{fig:ground_state_2d_n1}(a) and (b), such pairs of
points have particularly large values of $I_{ij}$, with the
largest occurring for pairs of points near the corners of the
Fermi surface, i.e., the pairs $\boldsymbol{k} = (\pm \delta k_x, \pi)$ and 
$\tilde{\boldsymbol{k}} = (\pi\pm\delta k_x, 0)$, with $\delta k_x = \pi / L_x$.
Note that there is asymmetry in the $k_x$--$k_y$ plane due to the mixed
boundary conditions (periodic in $y$ and antiperiodic in $x$).
The single-site entropy [Fig.~\ref{fig:ground_state_2d_n1}(d)]
reflects this structure in that the largest values occur at the
$\boldsymbol{k} = (\pm \pi,0)$ and $\boldsymbol{k} = (0,\pm \pi)$
corners of the Brillouin zone.
As can be seen in Fig.~\ref{fig:ground_state_2d_n1}(a), $I_{ij}$ has
a fairly diagonal structure for the optimal ordering, except for a
relatively limited region in the middle of the MPS site ordering.
However, this region 
is quite important for the convergence in that the DMRG steps with a
high number of states kept will be required in this region.

Doping the two-dimensional system deforms the Fermi surface and
destroys its perfect nesting, as is depicted in  
Fig.~\ref{fig:ground_state_2d_n075}(b).
The sites that correspond to those that showed the strongest correlations 
in the half-filled case are now less correlated.
Similarly to the one-dimensional case, reordering the MPS sites can 
significantly reduce the entanglement in the system, but a
``perfect'' ordering is again not possible because of
loops in the entanglement structure, which can be seen in
Figs.~\ref{fig:ground_state_2d_n1}(b) and
\ref{fig:ground_state_2d_n075}(b).
The site occupancy [Fig.~\ref{fig:ground_state_2d_n075}(c)] shows a
sharp jump at the Fermi surface, and the single-site entropy
[Fig.~\ref{fig:ground_state_2d_n075}(c)] a relatively broad peak, with
the highest values occurring at the corners of the Fermi surface, in
the $(k_x,0)$ and $(0,k_y)$ directions.

The mappings between MPS sites $i$ and momentum points 
$k \, / \, \boldsymbol{k}$ for optimal ordering for the
one-dimensional systems depicted in
Figs.~\ref{fig:ground_state_1d_n1} and \ref{fig:ground_state_1d_n075}
are listed in Table~\ref{tab:1d_ordering}, 
and the two-dimensional systems depicted in
Figs.~\ref{fig:ground_state_2d_n1} and \ref{fig:ground_state_2d_n075}
in Table~\ref{tab:2d_ordering}.
In the matrix plots of $I_{ij}$ [Figs.~\ref{fig:ground_state_1d_n1}(a),
\ref{fig:ground_state_1d_n075}(a), \ref{fig:ground_state_2d_n1}(a), 
and \ref{fig:ground_state_2d_n075}(a)], the first minor diagonals 
correspond to the correlation between neighboring sites,
the second minor diagonals to correlations between next-nearest
neighbors, etc.
Therefore, the task of optimizing the MPS ordering can be graphically
interpreted as permuting the columns and rows of $I_{ij}$ symmetrically
with the purpose of arranging its entries in descending order 
from the main diagonal $I_{ii}$ towards the outer edges $I_{1N}$ and $I_{N1}$.

\begin{figure}
\centering
\includegraphics[width=8.0cm]{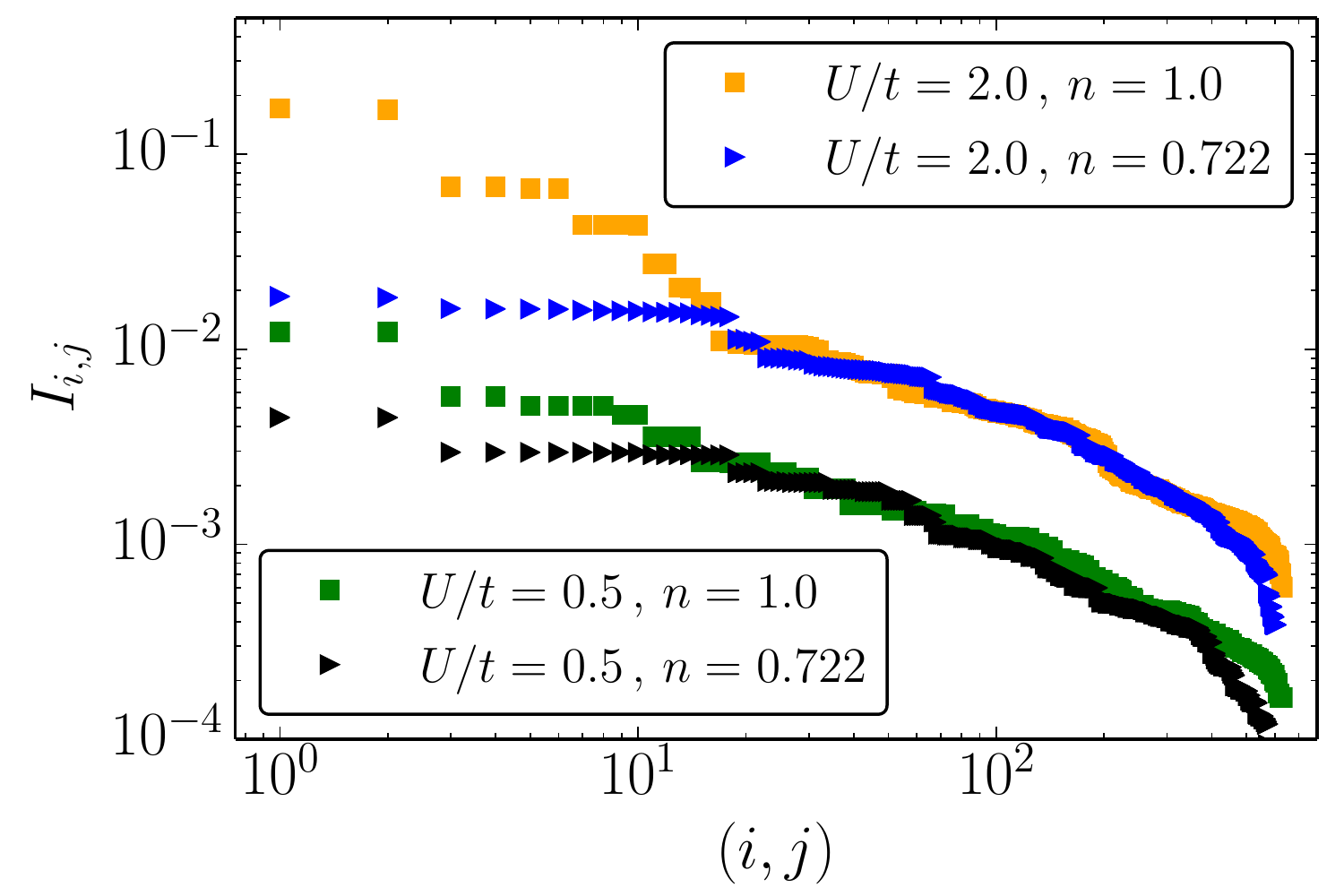}
\caption{ 
(Color online)
Two-site mutual information $I_{i,j}$ 
for the ground state of the $6{\times}6$ Hubbard model
in momentum space for different $U$ values and different fillings $n$. 
The values are sorted in descending order.
We include only unique entries of $I_{i,j}$ , i.e., take $i{>}j$.
}
\label{fig:mutual_information_scaling_2d}
\end{figure}

The decay rate of the elements of $I_{ij}$, provides
important information about the correlations in the 
system~\cite{Barcza-2014}.
This decay is depicted on a logarithmic scale in descending order for 
the $6{\times}6$ Hubbard model in 
Fig.~\ref{fig:mutual_information_scaling_2d}, for both the half-filled
and doped systems at two different $U$ values.
As can be seen, there are plateaus with jumps and changes in slope in
all four curves.
Adopting the terminology of
quantum chemistry, the small number of components with large $I_{ij}$
values are associated
with the so-called static correlations, while the large number of elements 
with small weight correspond to dynamic correlations. 
As the interaction strength $U$ is increased,
the curves shift upward for both the half-filled and doped cases,
indicating an increased level of entanglement. 
For the doped case, the static correlations are 
significantly smaller than for the half-filled case. 
This is due to the absence of the umklapp
processes in the doped case. 
Thus, we expect that the doped case can be studied more efficiently in
momentum-space representation than the half-filled case.
Note that in real space the opposite holds.  

We have also calculated the correlation patterns 
and the decay of the two-site mutual information
for  the $6{\times}6$ lattice at interaction strength $U=4.0$.
While the results do not reveal qualitatively new behavior,
the DMRG convergence becomes critical
due to the increased overall entanglement in the system.
Compared to $U=2.0$, the correlations between sites 
that fulfill the nesting condition in the half-filled case
become even more dominant.

\subsection{Wave-function coefficients}
\label{sec:wave_function_coefficients}

\begin{figure}
\centering
\includegraphics[width=8.0cm]{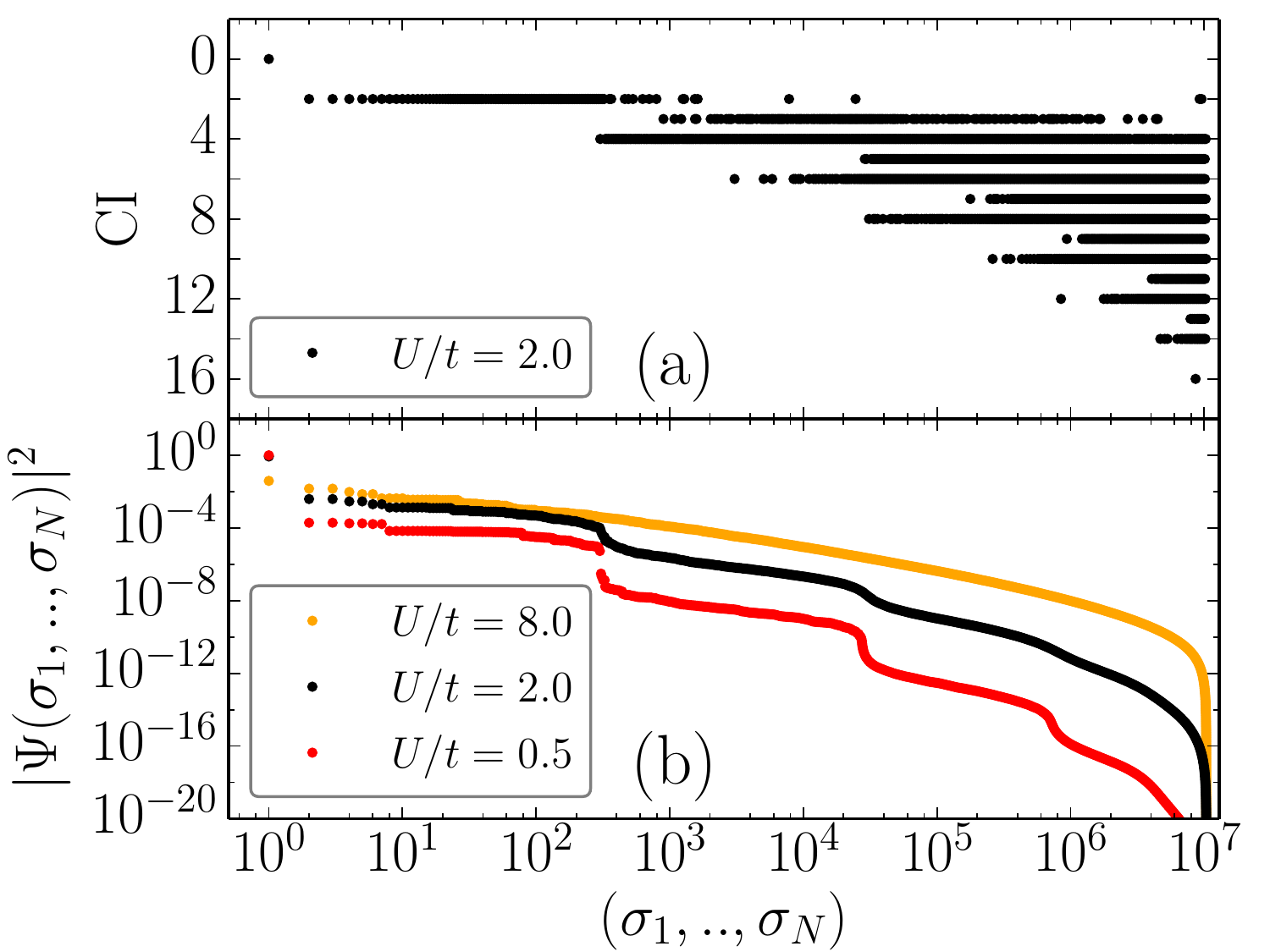}
\caption{
(Color online)
(a) Number of excitations from the Hartree-Fock state 
in the basis states $\left|\sigma_1,..,\sigma_N\right\rangle$
for $U=2.0$.
(b) Square of the coefficients $\Psi(\sigma_1,..,\sigma_N)$ of the ground state 
of the $4{\times}4$ Hubbard model in momentum space at half-filling
for different values of $U$.
The data in (a) and (b) is sorted by the value of 
$|\Psi(\sigma_1,..,\sigma_N)|^2$ in descending order.
}
\label{fig:ci_coefficients_2d}
\end{figure}

The wave-function coefficients $\Psi(\sigma_1,..,\sigma_N)$ 
in the full tensor representation 
$\Psi = \sum_{\{\sigma_1,..,\sigma_N\}} \Psi(\sigma_1,..,\sigma_N)
|\sigma_1,..,\sigma_N\rangle$
can also be calculated from the DMRG wave function within the MPS formalism.  
The squares of the coefficients $\Psi(\sigma_1,..,\sigma_N)$ of the
ground state of the half-filled $4{\times}4$ Hubbard model in momentum
space are shown in Fig.~\ref{fig:ci_coefficients_2d}(b) for different
values of $U$.
The increased level of entanglement is reflected by the 
increase in the weight of the wave function coefficients, a behavior
similar to that seen for the decay of $I_{ij}$ in
Fig.~\ref{fig:mutual_information_scaling_2d}. 
 
In order to make contact with the CI-expansion technique used in 
quantum chemistry, the number of excitations from the Hartree-Fock state ($\rm{CI}=0$) 
in the corresponding basis states $\left|\sigma_1,..,\sigma_N\right\rangle$
for $U=2.0$ are shown in Fig.~\ref{fig:ci_coefficients_2d}(a). 
The single CI ($\rm{CI}=1$) and the second highest ($\rm{CI}=N-1$) excitation 
are forbidden due to momentum conservation.
The most important excitations are the double, triple, and quadruple CI
determinants, but higher CI excitations also provide significant
contributions.
For small $U$ values, these excitations are reflected in the
behavior of the square of the coefficients of  
the wave function in that there are plateaus
with jumps.
The first few plateaus come mainly from the lower-order CI excitations.
This structure becomes smoother for larger interaction strength
and completely disappears for large $U$ values.
The decay of the coefficients with coefficient index also becomes
increasingly longer range with increasing $U$.
Therefore, a truncated
CI expansion would not converge sufficiently rapidly in the level of
excitation in the momentum-space representation due to strong
correlations in the system, except possibly at very small $U$ values.

\subsection{Block entropy and entanglement scaling} 
\label{sec:entropy_scaling}

\begin{figure}
\centering
\includegraphics[width=8.0cm]{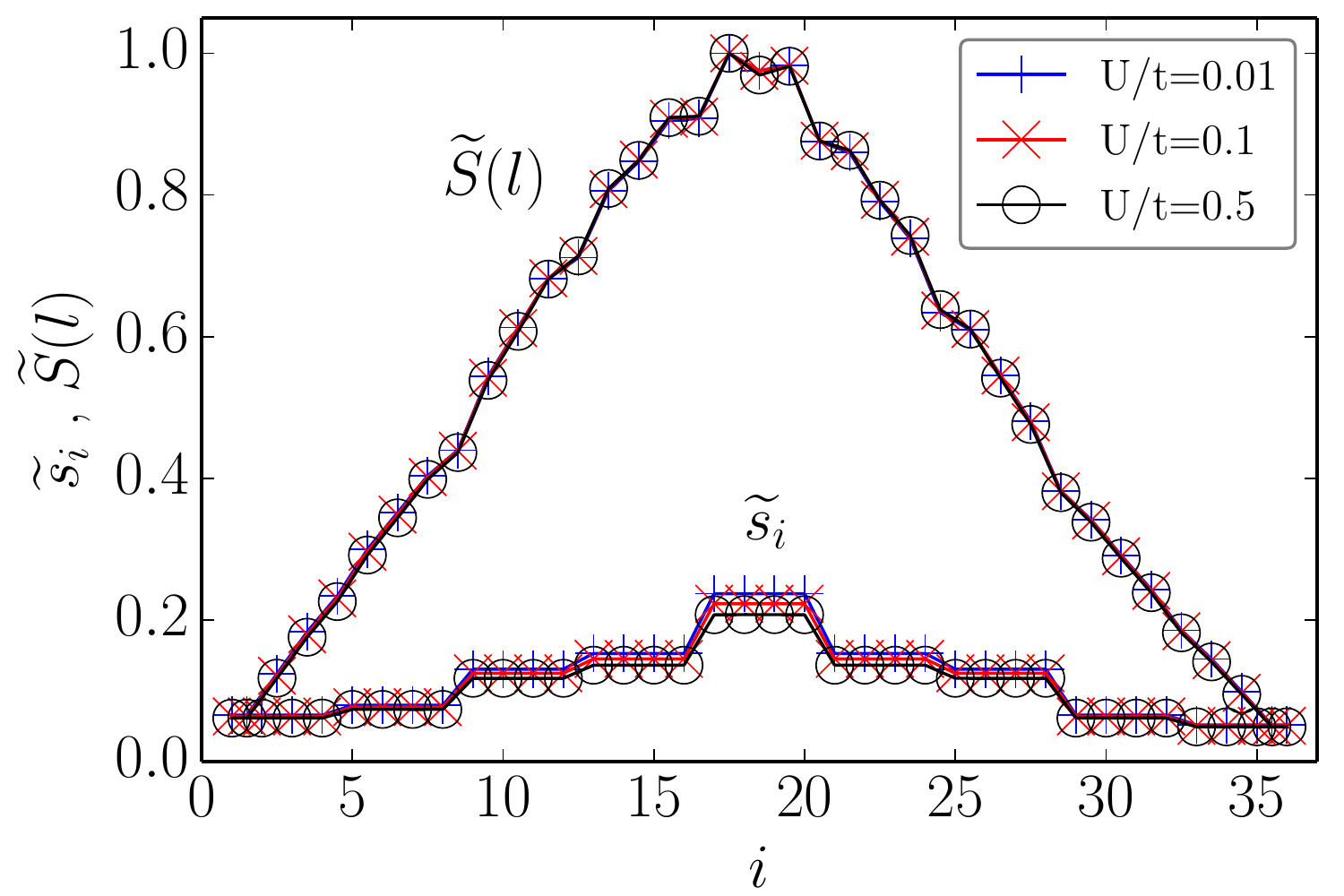}
\caption{ 
(Color online)
Normalized block entropy $\widetilde{S}(l)$ (upper curves) 
and single-site entropy $\widetilde{s}_i$ (lower curves)
of the ground state of the $6{\times}6$ Hubbard model in momentum space
for small values of $U$ at half-filling.
}
\label{fig:entropy_profiles_2d_6x6}
\end{figure}

After studying the entanglement patterns and correlations
for particular finite-sized systems, 
we now analyze the scaling of the entanglement as a function 
of interaction strength and system size.
We begin by examining the behavior of lattice-site-dependent
quantities at optimal ordering (i.e., as a function of the DMRG index
$i$) for the two-dimensional system, for which the behavior is not
known and is particularly interesting in view of the nontrivial
mapping of the two-dimensional momentum lattice to MPS sites.
In Fig.~\ref{fig:entropy_profiles_2d_6x6}, we display
the normalized block entropy 
$\widetilde{S}(l) \equiv S(l) / \mbox{max} [ S(l)]$  
and the similarly rescaled single-site entropy 
$\widetilde{s}_i \equiv s_{i} / \mbox{max} [ S(l) ]$
of the ground state of the half-filled $6{\times}6$ Hubbard model in
momentum space for small values of $U$.
As can be seen, the rescaled profiles of both quantities fall onto a
single curve to a good approximation, although the actual values of
both the block and the site entropies change significantly.

To understand the scaling behavior, we consider
the subadditivity of the entropy~\cite{Sagawa-2012}, i.e., 
the change in the block entropy in each 
 full step of the DMRG, in which an enlarged block  
with $l+1$ sites is formed from a block with $l$ sites and the 
$(l+1)^{\rm th}$ site~\cite{Legeza-2004}.
This entropy reduction is due to the correlations between the block
and the site and is given by the mutual information
\begin{eqnarray}
I(l) = S(l)+s_{l+1}-S(l+1)\,.
\label{eq:I-l}
\end{eqnarray}
In the DMRG, the set of block states is 
chosen by taking the eigenstates 
of the reduced density matrix with the largest eigenvalues.
It can be shown that this operation can be implemented as a
LOCC (local operation and classical communication) 
so that it cannot lead to an increase in
entanglement.
With no truncation, it holds that $I_{\rm tot}\equiv\sum_l
I(l)$ for a full sweep 
is equal to the sum of the single-site entropies,
$\sum_i s_i$. 
Thus, $I_{\rm tot}$ 
quantifies the total quantum
information encoded in the wave function. 
Therefore, the scaling of the site entropy as a function of $U$ 
and system size also determines the scaling of the block entropy. 
We will come back to the detailed scaling for weak $U$ once we
have investigated the system-size and $U$ dependence of the maximum
of the block entropy, $\max[S(l)]$.

\begin{figure}
\centering
\includegraphics[width=8.0cm]{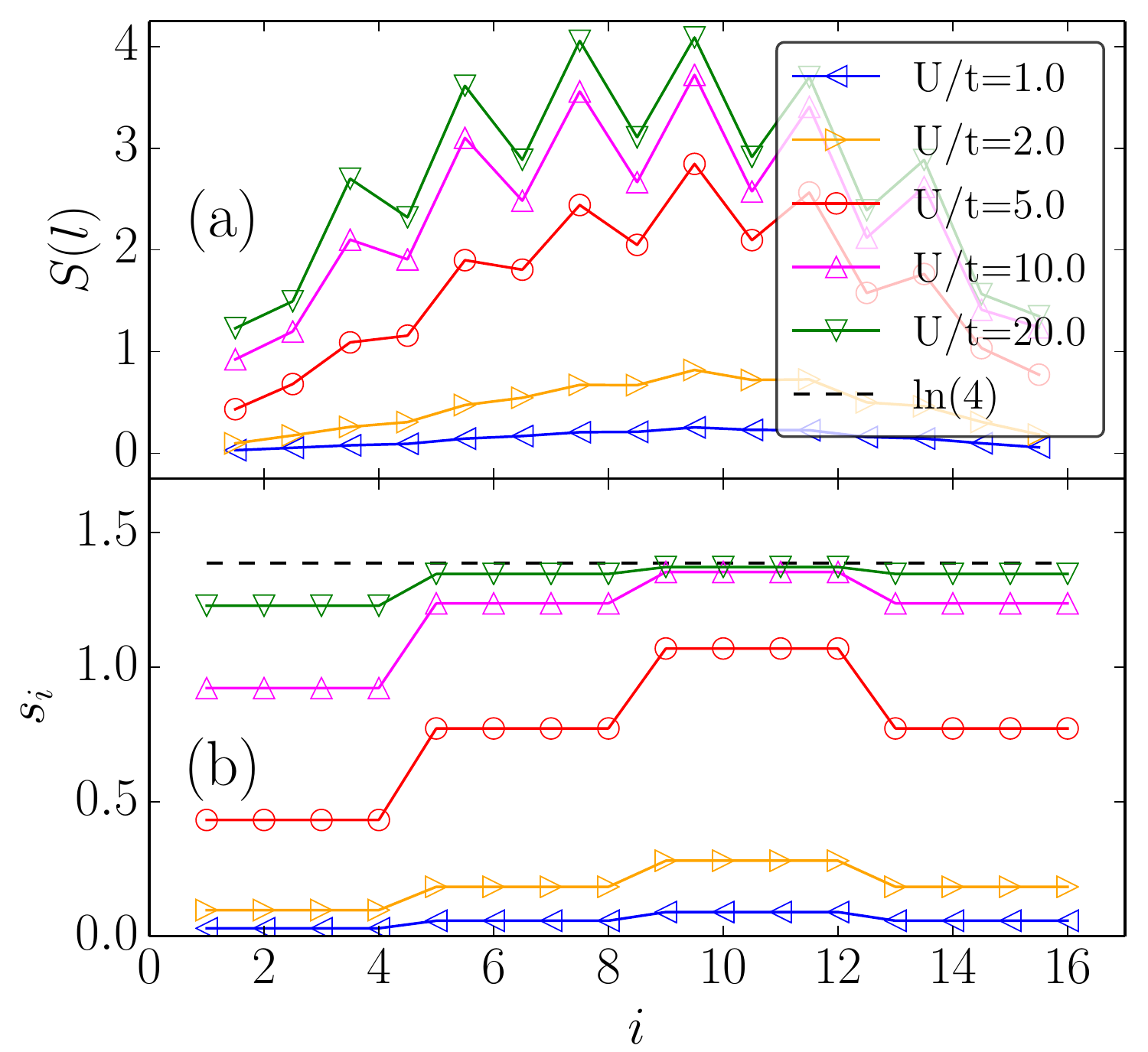}
\caption{ 
(Color online)
Block entropy $S(l)$ (a) and single-site entropy $s_i$ (b)
of the ground state of the $4{\times}4$ Hubbard model in momentum space 
for large values of $U$ at half-filling.
}
\label{fig:entropy_profiles_2d_4x4}
\end{figure}

Next, we examine the large-$U$ behavior of
the single-site and block entropies, depicted
in Fig.~\ref{fig:entropy_profiles_2d_4x4}.
In this case, not only the amplitude
but also the shape of the profiles changes; 
thus, we do not rescale the data.
As can be seen, 
above a critical value of $U$, some of the site entropies reach the 
maximum value of $\ln 4$, and the site entropy profile starts to broaden.
(In the $U\rightarrow\infty$ limit, all sites will have
$s_i=\ln 4$.)
Since the growth of the block entropy per site is bounded by
the single-site entropy due to Eq.~(\ref{eq:I-l}),
the block entropy profile and its maximum must also saturate.
Furthermore, in the half-filled case for these larger-$U$ values,
zig-zag peaks appear superimposed upon the profile.
These peaks correspond to those bipartite partitions of the system
that cut the system between two sites that fulfill the nesting
condition (which are the strongest correlated sites, as reflected in $I_{ij}$).
Therefore, the peaks are related to the zig-zag pattern seen in the
first minor diagonals of $I_{ij}$ in
Fig.~\ref{fig:ground_state_2d_n1}(a).

\begin{figure}
\centering
\includegraphics[width=8.0cm]{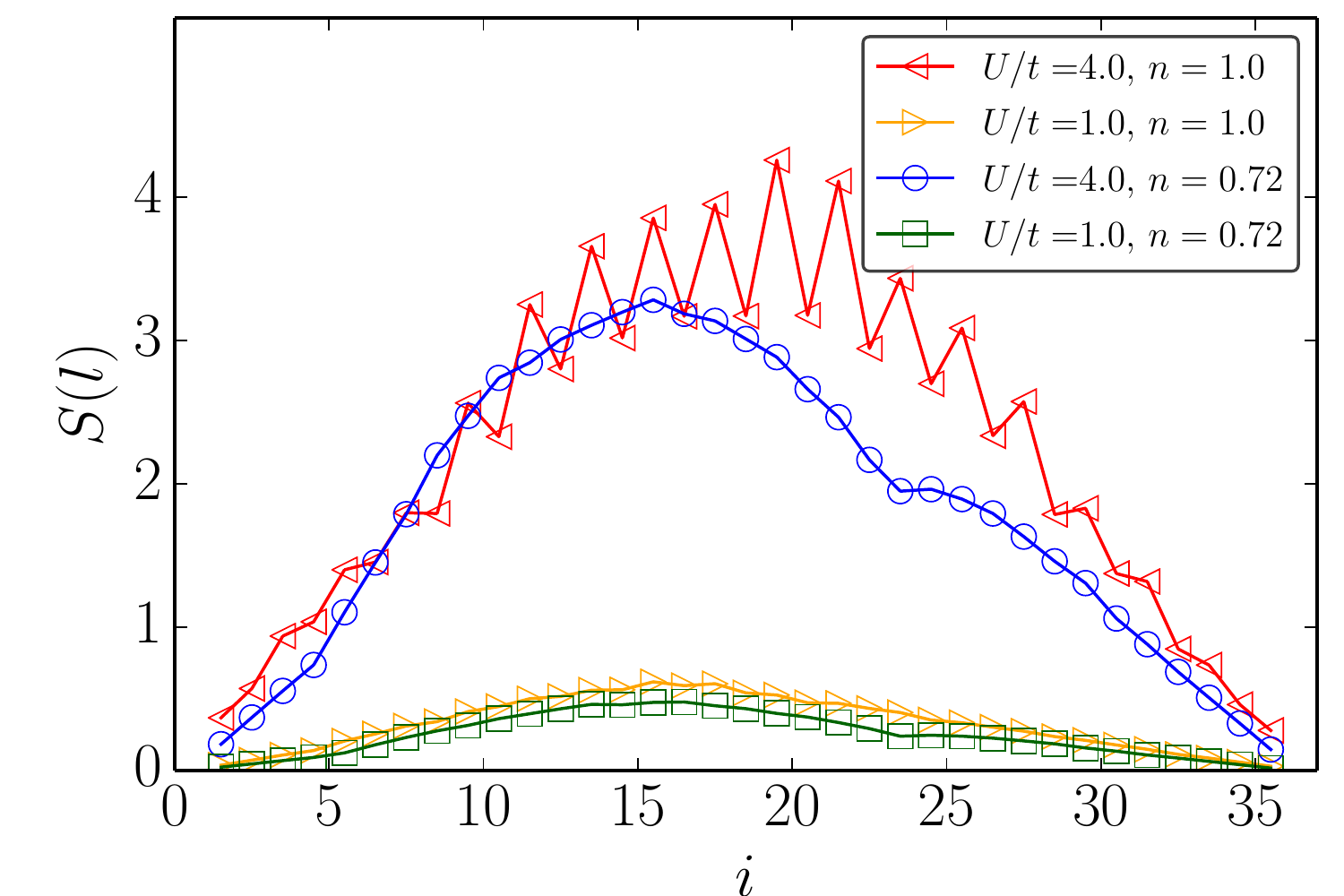}
\caption{
(Color online)
Block entropy profiles $S(l)$ of the $6{\times}6$
Hubbard model in momentum space for weak coupling $U/t=1.0$ and 
intermediate coupling $U/t=4.0$ for the half-filled $n=1.0$ and
doped $n=0.722$ case.
Every profile is calculated using the optimal ordering
for the particular set of parameters $U$, $n$.
}
\label{fig:entropy_profiles_2d_6x6_b}
\end{figure}

We now turn to the effect of doping.
In Fig.~\ref{fig:entropy_profiles_2d_6x6_b}, we display  
a comparison of block entropy profiles for the 
half-filled and doped system for weak and intermediate coupling
strength.
At $U=1$, the profiles are similar, with no additional structures
present.
At $U=4$, it is evident that the additional oscillating structures are
only present at half-filling.
This is due to the fact that the perfect nesting of the
Fermi surface  is destroyed in the doped system, so that the
corresponding bonds are less entangled, as discussed before, and no
additional peaks develop.
Note that for each parameter set, the block entropy is calculated for the 
corresponding optimal ordering and thus the MPS site indices $i$ 
do not correspond to the same $\boldsymbol{k}$~points.

\begin{figure}
\centering
\includegraphics[width=8.0cm]{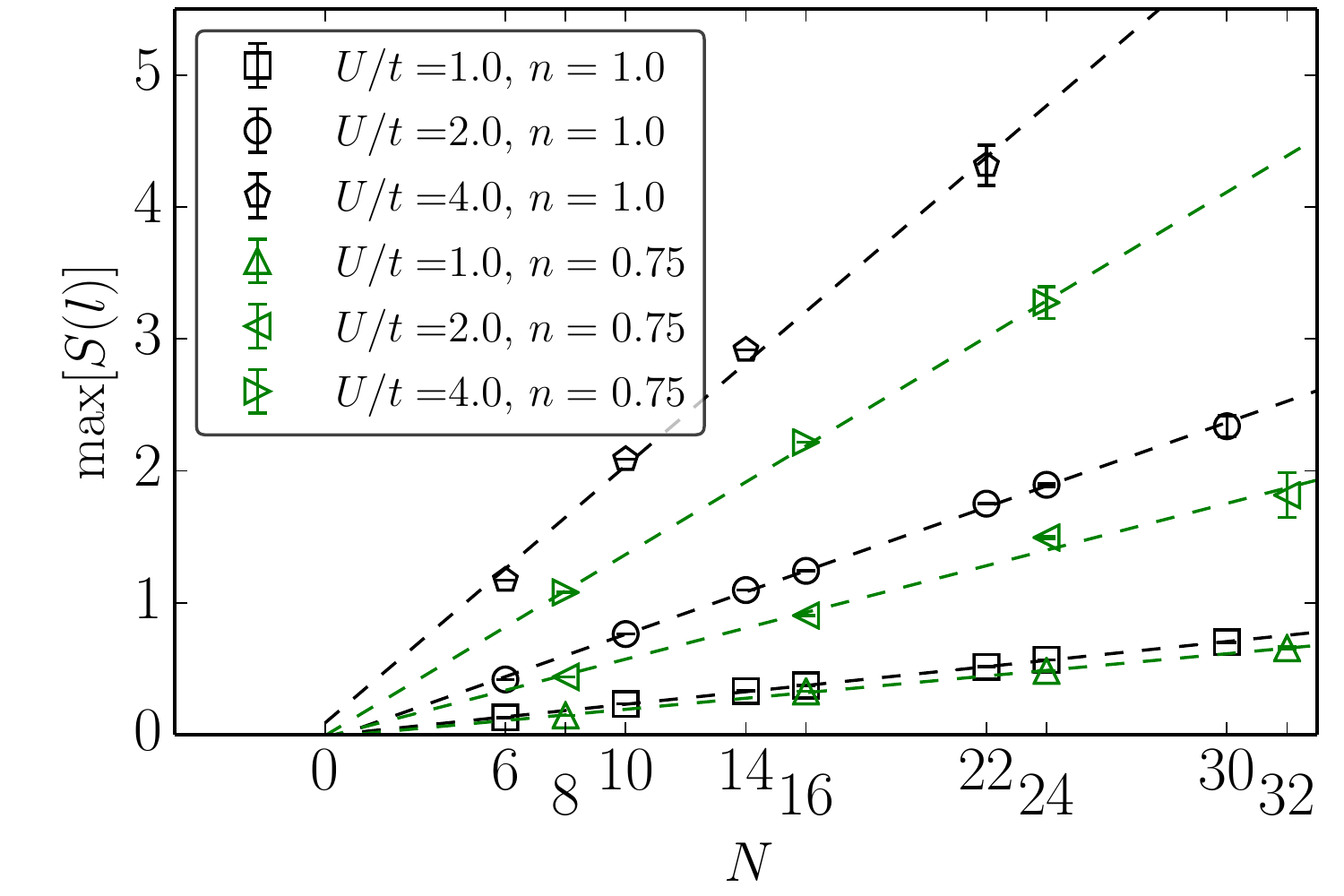}
\caption{
(Color online)
Maximum of the block entropy, $\max[S(l)]$,
as a function of the system volume $N$
for different values of $U$ 
for the one-dimensional Hubbard model 
in momentum space at half-filling and at three-eighths filling, 
$n = 0.75$. 
The dashed lines are linear fits to the data.
}
\label{fig:entropy_scaling_n_1d}
\end{figure}

We now analyze the entanglement scaling as a function of system size $N$,
starting with the one-dimensional model and then moving to the
two-dimensional case.
In both cases, the data points in the figures refer to
the maximum of the block entropy
taken from the last half sweep of each calculation, 
while the error bars are given by the difference
between these values and the zero-truncation extrapolated values.
(The extrapolated values are obtained
by repeatedly measuring the maximum of the block entropy during the sweeping process
while simultaneously increasing the number of block states kept and
thus lowering the truncation error.)
This procedure for error estimation generally tends to 
overestimate the actual error.
Figure~\ref{fig:entropy_scaling_n_1d} depicts 
the maximum of the block entropy, $\max[S(l)]$, for the half-filled, $n=1.0$, 
and for the doped, $n=0.75$, one-dimensional
Hubbard model as a function of the number of sites $N$
for various values of $U$.
The dashed lines, which are linear fits to the data,
show that the maximum of the block entropy scales with the
volume of the system. 
This scaling with volume for a wide range of $U$ values clearly shows
that the entropy area law does not hold for this case.
Interestingly, in momentum space,
the block entropy is appreciably smaller for the doped system
than for the half-filled system of the same size and 
interaction strength.
Consequently, the doped system is easier to treat computationally than
the half-filled system, in contrast to real space, in which the doped
case is generally harder to treat numerically.

\begin{figure}
\centering
\includegraphics[width=8.0cm]{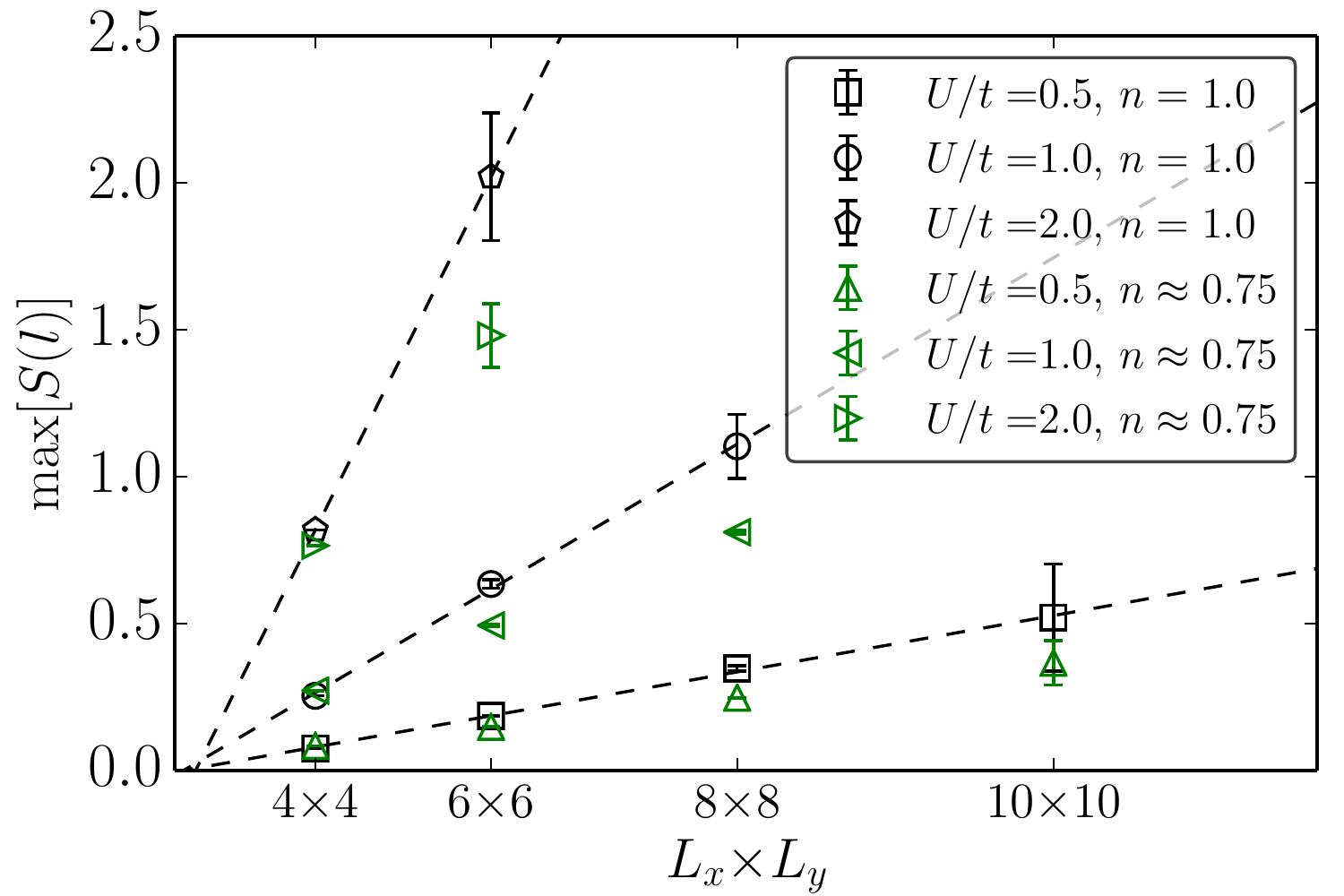}
\caption{
(Color online)
Maximum of the block entropy, $\max[S(l)]$,
as a function of the system volume $N=L_x L_y$
for different values of $U$ 
for the two dimensional Hubbard model 
in momentum space at half-filling and near $n=0.75$ filling 
($n=0.75$ for $4{\times}4$, $n=0.722$ for $6{\times}6$, 
$n=0.75$ for $8{\times}8$, $n=0.74$ for $10{\times}10$). 
The dashed lines are linear fits to the data at half-filling.
}
\label{fig:entropy_scaling_n_2d}
\end{figure}

For the two-dimensional system, Fig.~\ref{fig:entropy_scaling_n_2d} 
depicts the scaling of the maximum of the block entropy 
with system size for different values of $U$
for the half-filled, $n=1.0$, and for the doped, $n\approx0.75$, systems.
The dashed lines are linear fits to the data for the half-filled system,
indicating almost perfect scaling with the system size.
For the doped case, the block entropy for a particular value of $U$
is again noticeably smaller than for the half-filled system.
As we have seen earlier in the entropy profiles,
this effect becomes stronger for larger values of $U$.
Note that for the $4{\times}4$ system,
the nesting of the Fermi surface is not destroyed in the doped case;
therefore, the block entropies for the half-filled and doped cases 
do not differ much.

Regarding the question of the origin of the volume scaling of the
maximum block entropy, we argue as follows:
Within the momentum-space picture, what is important is how
strongly a given region in momentum space is correlated with other 
regions, especially regions that correspond
to DMRG blocks at optimal site ordering.
As the system size is increased, the density of the $\boldsymbol{k}$~points 
in each region of momentum space increases linearly with system size,
ultimately leading to a linear increase in the total correlation.
Furthermore, the correlation patterns in momentum space depicted in 
Figs.~\ref{fig:ground_state_1d_n1}(b), 
\ref{fig:ground_state_1d_n075}(b),
\ref{fig:ground_state_2d_n1}(b), and \ref{fig:ground_state_2d_n075}(b)
show that strong correlations occur between regions
close to each other as well as widely separated from each other.
In particular, correlation bonds of comparable strength
form loops, which cannot be disentangled globally
when the momentum-space lattice is mapped onto the sites of the MPS.
Thus, even for optimal site ordering,
regions of sites separated by a macroscopic number of sites in the MPS lattice,
i.e, a separation of up to an  appreciable fraction of the system size, 
will still be correlated with one another.
The growth of the total correlation with the site density,
in combination with the fact that the corrections are long range in the MPS,
finally leads to a linear scaling of the block entropy $S(l)$ 
with system size for all decompositions
and to volume-law scaling of the maximum block entropy.
The linear growth of the total correlation with subsystem size is also
reflected in the approximately triangular-shaped block entropy profiles in
Figs.~\ref{fig:entropy_profiles_2d_6x6} and
\ref{fig:entropy_profiles_2d_6x6_b}.
Note that reordering the system in an optimal way is still crucial
in order to obtain good convergence, even though it cannot avoid the
volume scaling of the entropy in the momentum-space Hubbard model.

\begin{figure}
\centering
\includegraphics[width=8.0cm]{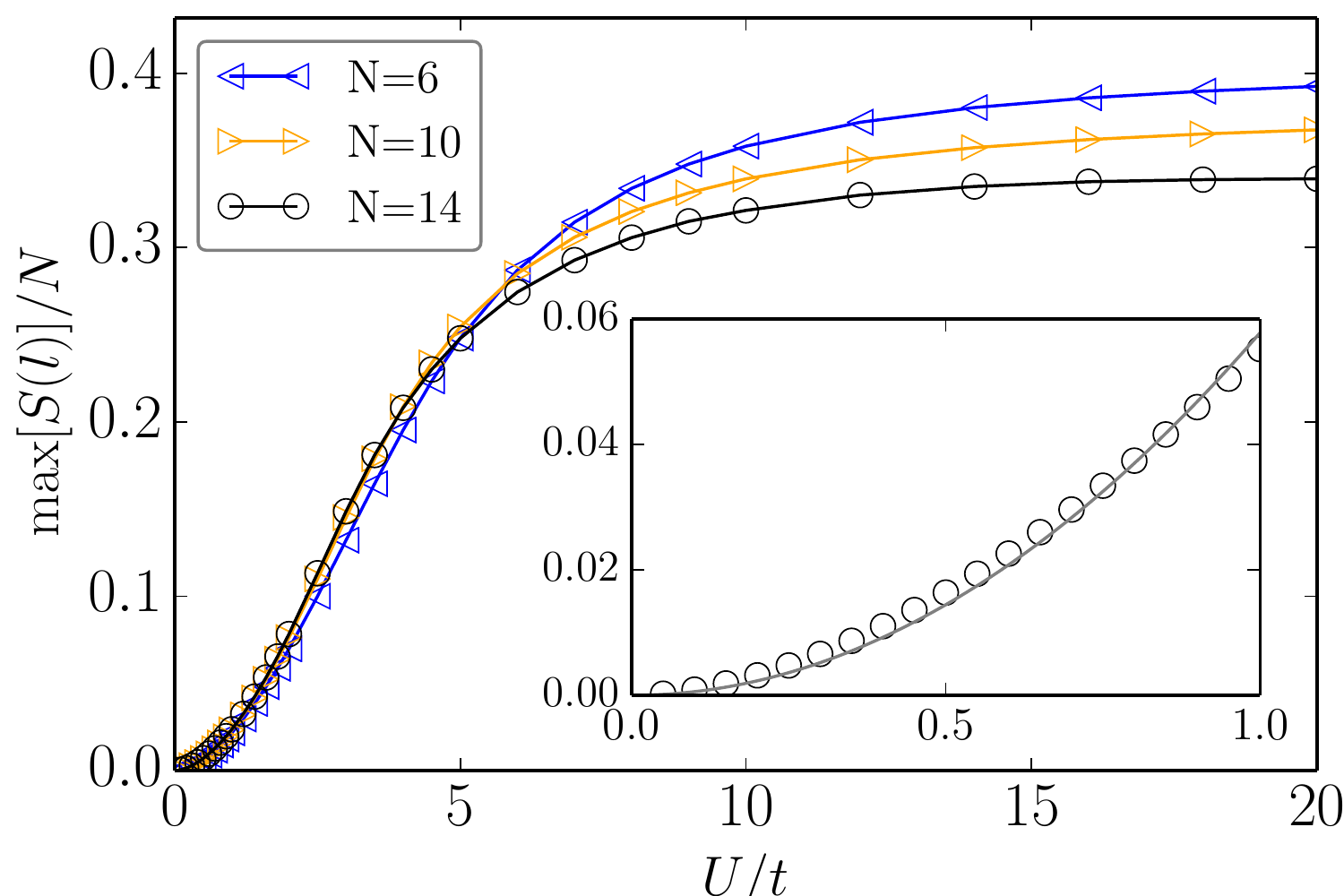}
\caption{
(Color online)
Maximum of the block entropy $S(l)$ per site as a function of $U$
calculated exactly for the $N=6,10,14$ site 
Hubbard model in momentum space at half-filling.
The inset shows a quadratic fit (gray line) for $0 \leq U \leq 1$ and $N=14$.
}
\label{fig:entropy_scaling_u_1d}
\end{figure}

Having established the volume scaling of the block
entropy, we finally investigate the $U$ dependence of the volume-scaled
maximum block entropy, taking the
half-filled case. 
We divide out the volume scaling in $\max[S(l)]$ by taking 
$\max[S(l)]/N$ so that we can compare the behavior for different
system sizes.
We start with the one-dimensional case, depicted in
Fig.~\ref{fig:entropy_scaling_u_1d}.
For $U \rightarrow 0$, the block entropy must 
vanish because
the noninteracting ground state is a product state in momentum space.
For weak coupling, the maximum block entropy scales with $U^2$,
as can be seen in the inset, where a fit to $U^2$ is shown.
For large $U$, the maximum of the block entropy saturates for a given
system size, but the upper limit seems to be significantly smaller
than its theoretical limit of $\ln 2$.
Furthermore, while the curves for different system sizes 
fall on top of each other (as expected given a volume law for the
system size scaling) for small $U$, for large $U$ the limiting value
decreases somewhat with system size,
indicating a sub-linear scaling.
However, only relatively small system sizes are accessible in the
large-$U$ limit, so that a definitive analysis of this effect cannot
be carried out.

\begin{figure}
\centering
\includegraphics[width=8.0cm]{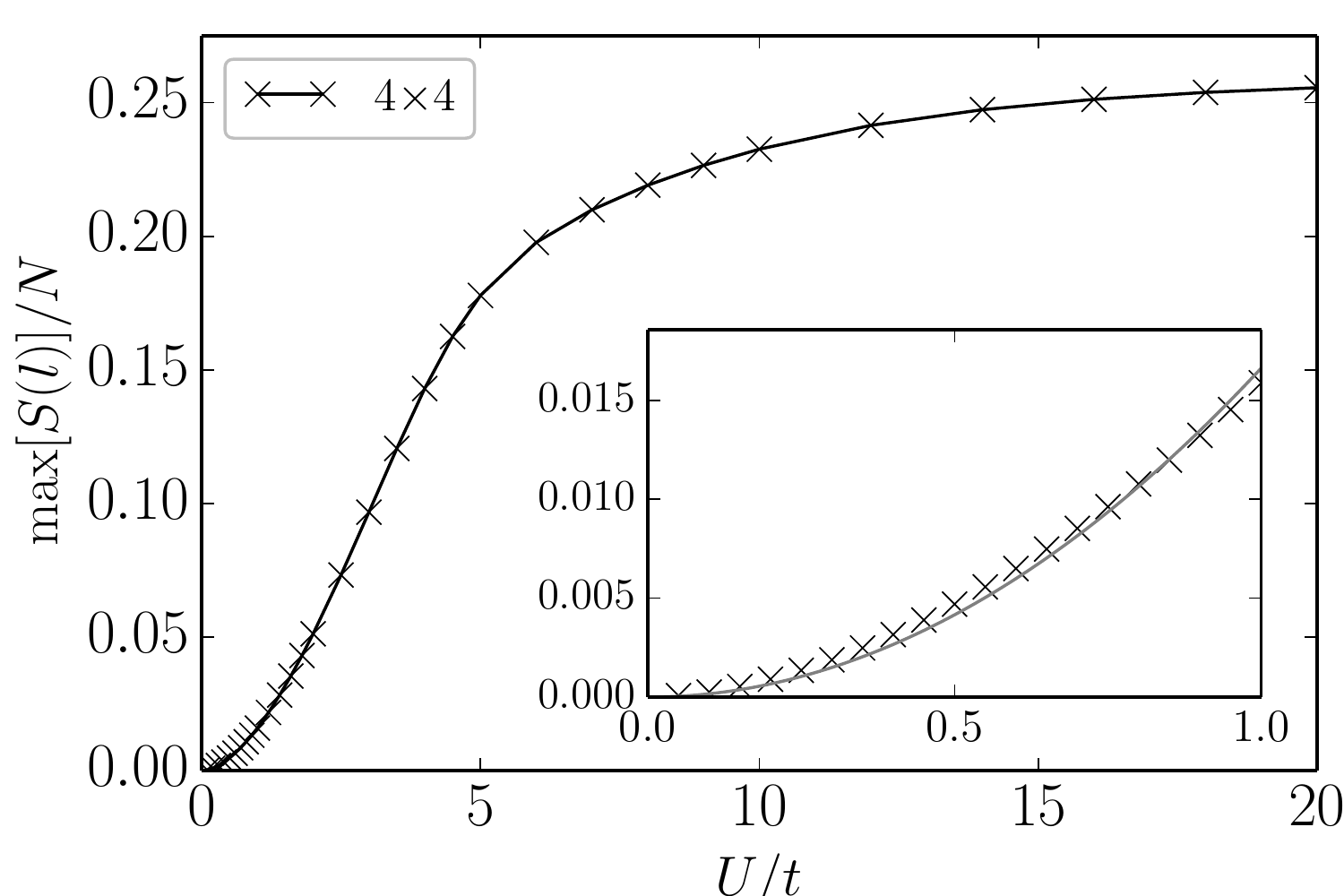}
\caption{
Maximum of the block entropy $S(l)$ per site as a function of $U$
calculated exactly for the $4{\times}4$
Hubbard model in momentum space at half-filling.
The inset shows a quadratic fit (gray line) for $0 \leq U \leq 1$.
}
\label{fig:entropy_scaling_u_2d}
\end{figure}

For the two-dimensional case, an analysis for the complete range of
$U$ values can only be carried out for the exactly treatable $4{\times}
4$ system.
For this system, the behavior of the maximum block entropy 
as a function of $U$, shown in Fig.~\ref{fig:entropy_scaling_u_2d}, is
similar to that in one dimension, 
with quadratic scaling in $U$ in the weak-coupling regime and saturation
of the maximum of the block entropy in the strong-coupling limit.
We have previously shown that the block entropy
$S(l)$ and the single-site entropy $s_i$ show the
same scaling behavior as a function of $U$ for all $l$ or $i$,
respectively.
Therefore, $S(l)$ and $s_{i}$ also scale quadratically with $U$ at
weak coupling.
Note that the saturation value for large $U$ is actually significantly
smaller in the two-dimensional case than in the one-dimensional case.

\section{Discussion and Conclusion}
\label{sec:conclusion}

In this paper, we have applied an efficient, optimized version of the
momentum-space DMRG to the Hubbard model in one and two dimensions.
The effectiveness of our method has allowed us to carry out an extensive
study of the entanglement structure for both the one-dimensional and
two-dimensional models at half-filling and at moderate doping.
In particular, we have examined a number of quantum-information-based
measures in order to understand the entanglement and pair-wise
correlation of sites in the single-particle momentum basis.

By analyzing the correlations between individual sites
in terms of the two-site mutual information,
we have been able to determine what type of excitations of the
noninteracting system
are most important within the interacting ground state.
For the one-dimensional half-filled system,
we have found that the strongest correlations are between 
pairs of sites that are close to the Fermi points but are separated by
$\pi$ in momentum space. 
Excitations over these sites, due to their relative positioning in momentum space,
are energetically favorable and have a momentum transfer
of $2\pi$, which is necessary for umklapp scattering processes.
In the doped case, the pairs of sites close to the now shifted Fermi
points are no longer separated by $\pi$, and therefore umklapp scattering 
and the corresponding excitations are forbidden by momentum conservation.
With fewer favorable excitations allowed in the ground-state space,
the correlations drop significantly,
leading to lower block entanglement for the corresponding bipartite
decompositions of the system.
For the two-dimensional system, the perfect nesting of the Fermi
surface in the half-filled case is responsible for an analogous effect.
Pairs of sites near the Fermi surface that are 
separated in the $\boldsymbol{k}$ plane by $\left(\pi,\pi\right)$
thus allow for a larger number of energetically low-lying excitations.
When the system is doped, the deformation of the Fermi surface
destroys the nesting and truncates the number of 
energetically favorable excitations allowed 
in the ground-state space.
Apart from these observations,
we have used the 
results directly to obtain an optimal 
ordering of momentum sites within the MPS chain for each situation
in order to significantly improve the DMRG convergence.

In a second step, we have investigated the behavior of the von Neumann
entropies, in particular, the single-site entropy and 
the subsystem entropy of the DMRG blocks.
The most important indicator for the computational costs and convergence
of the DMRG  is the scaling of the maximum of the block
entropy, which we have analyzed as a function of the system volume and
coupling strength.
For the half-filled and doped systems we have found that the maximum
block entropy scales proportionally to the number of sites.
We understand this from the two-site mutual information measurements,
which indicate that there is an increase in total correlation
between regions in momentum space proportional to the density of
momentum points and that the correlations in our system are long range
in the MPS structure in spite of the optimization of the site ordering.
The combination of these two effects leads to volume-law scaling of
the block entropy.

Once the volume dependence is factored out, we obtain almost universal
curves for the maximum of the block entropy per site for both the one-
and two-dimensional Hubbard models, at least at half-filling.
These curves must go to zero at zero interaction and increase 
with the square of the interaction strength for weak coupling,
retaining a perfect volume scaling.
The presence of volume scaling of the entropy at arbitrary weak
interaction is a distinct indication of the fundamental limitations of
perturbative approaches for the Hubbard model in one and two dimensions.
For strong coupling, the maximum entropy per site saturates,
with a boundary value lying clearly below its theoretical limit and
with a volume scaling that seems to be slightly sublinear, at least in
the one-dimensional, half-filled case.
We find this result somewhat surprising and slightly less unfavorable
than naively expected, even though, admittedly, the high value of the
maximum entropy per site and the approximate volume law preclude any
well-controlled MPS--based approaches.

Despite these limitations, the momentum-space DMRG also has
some interesting favorable properties.  
While the convergence of the real-space DMRG
becomes worse for doped systems, we have found that, in momentum space,
the block entanglement decreases and the convergence improves
upon doping.  
In addition, our highly optimized and parallelized code has enabled us
to keep a comparably large number of block states at reasonable
computational cost, a number of states that is typically significantly
larger than a comparably optimized real-space DMRG code for similar
system size and parameters.
Since the coefficient of the volume scaling can be made arbitrarily
small at weak interaction, the k-DMRG could nevertheless yield
variationally competitive results, at least relative to the real-space
DMRG, for particular moderate system sizes at sufficiently weak interaction.

One possibility to overcome the main problems of momentum-space
DMRG while keeping some of its advantages is to use a hybrid 
real- and momentum-space representation, i.e., take a momentum basis
in the transverse direction and a real-space basis in the other
direction.
In this way, the translational invariance and good momentum quantum
numbers are preserved in the transverse direction, but the volume-law
entanglement due to long-range interactions the
in the longitudinal direction can be avoided.
  
\acknowledgements{
We thank S.~R.~White, D.~J.~Scalapino, G.~Barcza, S.~R.~Manmana, P.~R.~Corboz,  and Sz.~Szalay for
useful discussions.
This work was supported in part by
the Deutsche
Forschungsgemeinschaft (DFG) through Grant
No.~NO~314/5-1 in Research Unit FOR 1807 and the 
Hungarian Research Fund (OTKA) through Grants No.~K100908 and No.~NN110360.
}

\end{document}